\documentclass[final,12pt,authoryear,times]{elsarticle}
\usepackage[utf8]{inputenc}
\usepackage{amsmath}
\usepackage{mathrsfs}
\usepackage{graphicx}
\usepackage{epstopdf}
\usepackage{enumitem}
\usepackage{siunitx}
\usepackage{float}
\usepackage{mathtools}
\usepackage{xcolor}
\usepackage{subcaption}
\usepackage{algorithm}
\usepackage{algorithmic}
\usepackage{textcomp} 
\newcommand{\comment}[1]{} 

\setlist{leftmargin=5cm}
 \usepackage[modulo]{lineno}

 \usepackage{etoolbox}

\makeatletter   \def\ps@pprintTitle{%
      \let\@oddhead\@empty
      \let\@evenhead\@empty
         \def\@oddfoot{\centerline{\thepage}}%
      \let\@evenfoot\@oddfoot
   }
\makeatother

\begin{document}

\title{Periodic necking of misfit hyperelastic filaments embedded in a soft matrix }

\author[csu,cee]{Jian Li}
\author[me]{Hannah Varner}
\author[cee,me]{Tal Cohen\corref{cor1}}
\ead{talco@mit.edu}

\address[csu]{Key Laboratory of Traffic Safety on Track of Ministry of Education, School of Traffic and Transportation Engineering, Central South University, Changsha 410075, China}

\address[cee]{Department of Civil and Environmental Engineering, Massachusetts Institute of Technology, Cambridge, MA 02139}
\address[me]{Department of Mechanical Engineering, Massachusetts Institute of Technology, Cambridge, MA 02139}

\cortext[cor1]{Corresponding author.}

\begin{abstract}
The necking instability is a precursor to  tensile failure and rupture of materials. A quasistatically loaded free-standing uniaxial specimen  typically exhibits necking at a single location, thus corresponding to a long wavelength  bifurcation mode. If confined to a substrate or embedded in a matrix the same filament can exhibit periodic necking and fragmentation thus creating segments of finite length. While such periodic  instabilities have been extensively studied in ductile metal filaments and thin sheets, less is known about necking in hyperelastic materials. Nonetheless, in recent years, there has been a renewed interest in the role of necking in novel materials, for  the advancement of fabrication processes and to explain fragmentation phenomena observed in 3D printed active biological matter. In both cases materials are not well described by the existing frameworks that employ $J_2$ deformation theory plasticity, and existing studies do not account for the significant role of misfit stretches that may emerge in these systems through chemical, or biological contraction. To address these limitations, in this paper we begin by experimentally demonstrating the role of the surrounding matrix on the necking and fragmentation of a compliant filament embedded in a tunable rubber matrix.
Using a generalized hyperelastic model with strain softening, our analytical bifurcation analysis explains the experimental observations and is shown to agree with numerical predictions. The analysis reveals three distinct bifurcation modes: the long wavelength necking, thus recovering the Considère criterion; the periodic necking observed in our experiments; and a short wavelength mode that is characterized  by localization along the center cord of the filament and is   independent of the film-to-matrix stiffness ratio. We find that the softening coefficient and the filament misfit stretch can significantly influence the stability threshold and observed wavelength, respectively. 
 Our results can guide the design and fabrication of novel composite materials  and  can explain the fragmentation processes observed in active biological materials.

\end{abstract}
\maketitle
\nolinenumbers
\section{Introduction}

Materials composed of thin filaments\footnote{In this manuscript we will alternate between the terms `filament', and `fiber' to refer to slender rod like structures. The terms `film', and 'sheet' will be used to describe the plane-strain equivalent. } that are embedded in a matrix are ubiquitous in both nature and in synthetic systems. Examples of the former include microtubules in the living cell \citep{caporizzo2022microtubule,matis2020mechanical}, axons in brain tissue \citep{JOHNSON201335}, and tendons \citep{zitnay2018load},  while the latter is  found in fiber reinforced composites, a common paradigm for design of high performance materials \citep{li2018instabilities,huang2019superior,cui2020fiber,kabir2020critical}. Depending on the material properties and loading conditions, such systems may exhibit various instability phenomena, e.g. buckling, necking, and fracturing  \citep{steif1986periodic,nardinocchi2017swelling,colin2019layer,cerik2020use,riccobelli2021active}. Elucidating the mechanics of these phenomena is crucial in many emerging applications, including flexible electronics \citep{yan2020thermally}, and the design of enhanced  toughening of materials \citep{cooper2019toughening} that employ multi-material architectures \citep{zhu2005fabrication}.   

When subjected to compressive loads beyond a critical value, an isolated filament undergoes the classical Euler buckling. That stability threshold and the resulting wavelength can be offset by the support of a confining matrix, as shown by \cite{herrmann1967response}, and  recently realized in 3D-printed deformable fiber composites by \cite{li2018instabilities}. Therein, explicit formulas are obtained and predict the critical strain and the buckled wavelength  for a single infinitely long fiber embedded in a soft matrix. A different study, by \cite{zhao2016buckling}, illustrates how the critical strain at onset of buckling of a single fiber of finite length can be tuned over several orders of magnitude by modifying the fiber length to substrate thickness ratio. Other studies  show that slender fibers can develop non-planar patterns in the post-buckling regime \citep{su2014buckling,chen2017helical}.

When a filament is subjected to tensile loads, it will not buckle; instead, it may exhibit a necking instability and ultimately it may rupture, depending on the constitutive response of the material. For ductile materials, the localization of deformation to a necking region emerges due to the competition between stress increase and area reduction.  \cite{Considerecrete1885} first derived the criteria for the onset of necking by  identifying the maximum load in the uni-axial  stress-strain behavior of the material. Since then, the necking instability has been studied extensively considering various materials and  loading scenarios. \cite{hill1952discontinuous} and later \cite{storen1975localized} developed  theoretical foundations for the prediction of necking in ductile metal sheets. Hyperelastic materials can also exhibit a necking instability and obey the same Considère condition. They will thus exhibit necking if a maximal stress exists in the stress-strain behavior, as explained by \cite{audoly2016analysis}, which provided also a comparison between necking in hyperelastic and elastoplastic materials. It should be noted that such stress softening is commonly observed in soft polymeric materials  \citep{chaudhuri2007reversible,lwin2022rigidity,burla2020connectivity}. Nonetheless, even in absence of a maximum stress in the uni-axial response,  
 \cite{lin2016fringe} showed that a theoretical necking instability can emerge in constrained hyperelastic layers of high aspect ratio that are pulled along their thin direction.   Additional studies  show that the nonlinear constitutive response \citep{leonov2002theory,morovati2020necking}, surface tension \citep{mora2010capillarity,yan2022an,li2022surface}, and compressibility \citep{dortdivanlioglu2022plateau},  can significantly affect the necking instability. 

For  films or filaments  that are either deposited on a compliant substrate, or fully embedded in it, the surrounding matrix can significantly affect the necking instability and the subsequent fracture. Several studies have considered the emergence of periodic necks in metal films on compliant substrates, and show that the film and substrate properties cooperatively determine the resulting crack patterns \citep{thouless2011periodic,douville2011fracture,li2006deformability,xue2007neck,xue2008neck, jia2019bifurcation}. For the specific material system, all of these studies employed the finite strain $J_2$ deformation theory  plasticity with weak hardening. This particular choice of constitutive model (as opposed to $J_2$ flow theory plasticity, for example) is argued in earlier studies to provide better representation of the experimental observations \citep{storen1975localized, hutchinson1978sheet}.

Beyond metals, in recent years a renewed  interest in necking of confined filaments has emerged with the advancement of methods for multi-material 3D fabrication \citep{cho2016engineering,li2019domain}, including 3D printing of active biological matter \citep{daly2021bioprinting}.   The necking and subsequent fragmentation of  polymer fibers via cold drawing   was considered by \cite{shabahang2016controlled} to create periodic segments of similar lengths, as an approach for mass-production of short length nano-fibers. Bio-printed structural elements exhibit similar fragmentation behavior \citep{morley2019quantitative,wang2013necking} that emerges spontaneously, without application of external loads. It is thought to  result from cell contraction, which induces a `misfit' stretch or `pre-stress' into the system. This is a phenomenon is commonly observed in biological tissue \citep{du2019prescribing,liu2020growth}, and  in synthetic materials due to mismatch in chemical and thermal shrinkage during polymer curing \citep{kravchenko2016chemical, saba2018review}.  An approach to mimic this  behaviour as a mechanism for toughening of soft materials was recently demonstrated by \cite{cooper2019toughening}, showing sequential fracturing of a Gallium core embedded in an elastomer. In contrast to the earlier studies that have focused on ductile metals, to explain and predict the observations in novel soft polymer composites and in biologically active materials, there is a need for a more comprehensive and systematic understanding of the role of the specific hyperelastic constitutive response and the pre-stress, on the necking instability. In this work we aim to fill this gap by considering a generalized hyperelastic constitutive model with varying degrees of strain-softening and we examine its effect on the necking of pre-stretched confined filaments. 

The the paper is organized as follows: Experimental observations are presented in  Section 2, and are followed by the mathematical formulation in Section 3. Our theory extends the classical  formulation of \cite{biot1963surface}, as elegantly revisited by \cite{cao2012wrinkles} and later employed by  \cite{holland2017instabilities}. We present a complete derivation of the governing equations for the homogeneous deformation and the perturbation approach, considering a generalized material model with strain-softening and varying levels of pre-stretch. Section 4 describes the details of the numerical instability and  post-buckling analyses by using the \textcolor{black}{Finite Element Method (FEM)}. Results are discussed in Section 5, and we offer some concluding remarks in Section 6.

\section{Experimental observations}

To demonstrate the role of the constitutive properties in determining the critical threshold for the necking and subsequent fragmentation of an embedded filament, we use soft PDMS (Polydimethylsiloxane) rubber matrix material (Sylgard 184, Dow Corning) with an embedded filament of commercial putty (OATY, \raisebox{\depth}{\#}920033-5). This filament material is chosen for its softening behavior in tension: a putty filament will exhibit localization and necking at one location when subjected to tensile loads in isolation (i.e. without a PDMS matrix), corresponding to an infinite wavelength. The shear modulus of the filament $\mu_f$ was found to be $74.9$~kPa via compression testing at a stretch rate of $1\times10^{-3} s^{-1}$.  The modulus of the matrix $\mu_m$ is tuned by altering the base:cross-linker mass ratio to be stiffer or softer than the filament to obtain $\mu_m$=312.5~kPa for 15:1 and $\mu_m$=48.3~kPa for 40:1. These values are measured by a tensile test conducted at a at stretch rate of $1\times10^{-3} s^{-1}$ for the stiffer material (as detailed in Appendix A), and from \cite{raayai2019volume} for the softer material.

Our samples are fabricated using a two-step process resulting in a final test specimen containing a 2~mm diameter cylindrical filament centered in the matrix `dog-bone’ with a gauge length of 85~mm, width of 18~mm, and height of 13~mm. To achieve this, the PDMS is combined using a planetary centrifugal mixer, and then poured to the half-way point of the mold (6.35~mm). In order to prevent settling of the filament, the sample is degassed in a vacuum and pre-cured for 10 minutes at 100 \textdegree C before placing the filament and filling the remainder of the mold. The test samples with the putty filament are cured in an oven for 2.5 hours at 100 \textdegree C and then cooled to room temperature before testing. Tensile experiments are performed using an Instron 5943 Universal Testing Machine at a constant stretch rate of $10^{-3} s^{-1}$. All tensile experiments are performed within two days of the fabrication. The deformation sequences are recorded by a DSLR camera.

\begin{figure}[H]
    \centering
    \includegraphics[width=1\textwidth]{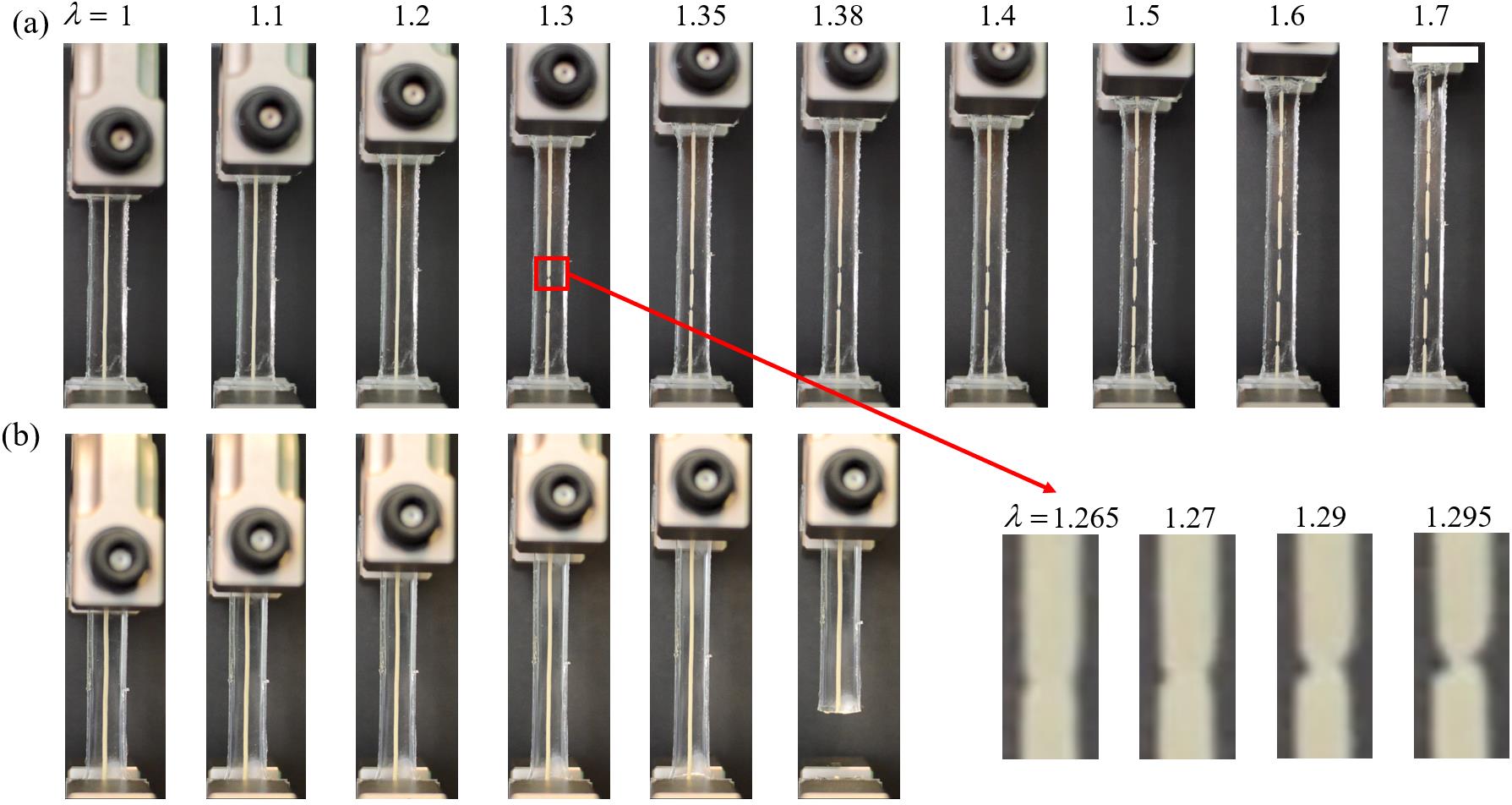} 
    \caption{\textbf{The stiffness of the matrix influences the onset of necking.} Images show the deformation sequences of circular filaments with  shear modulus $\mu_f=74.9$~kPa  embedded in PDMS matrix with different stiffnesses: (a) $\mu_m=48.3$~kPa,  and (b)  $\mu_m=312.5$~kPa.  Both samples are  subjected to  tensile deformation.  The recorded longitudinal stretch $\lambda$ (shown on the top) corresponds to both cases. 
    Enlarged images of the formation of the first neck in the sample with $\mu_m=48.3$~kPa are shown at higher resolution of applied stretch in the bottom right side of the figure. The development of necking instability in a free-standing circular putty film subjected to tensile deformation is illustrated in Appendix B. Scale bar: 30~mm. }
    \label{Fig.1}
\end{figure}

The experimental results of two representative tests are shown in Fig.~\ref{Fig.1}. We observe that the  matrix plays a crucial role in determining the necking and fragmentation of the film. For  a softer matrix in Fig.~\ref{Fig.1}(a) with film-to-matrix shear modulus ratio  $\mu_f/\mu_m\sim 1.6$, the PDMS matrix and film initially deform uniformly. Upon arriving at an applied stretch of $\lambda\sim 1.265$ (see inset in Fig. \ref{Fig.1}) the first indication of necking occurs, followed by elongation of the neck up until fragmentation and the formation of an intermediate segment,  as seen at the applied stretch of   $\lambda\sim 1.3$. With the further increase of the applied deformation, additional segments of similar lengths emerge. Overall, the average segment length in this sample is 12.1 mm, with a standard deviation of 2.23 mm (7 segments).
Note that the necking and fragmentation occurs sequentially. This is due to imperfection sensitivity, nonetheless the overall distribution of necks is expected to be dictated by the mechanical interaction between the filament and the matrix. Similar phenomena have  also been observed in glass fiber reinforced composites \citep{shabahang2016controlled} and in metal reinforced composites \citep{cooper2019toughening} subjected to tensile deformation along the fiber axes. 

For the stiffer matrix with  $\mu_f/\mu_m\sim 0.24$, the film and matrix deform uniformly until  fracture of the matrix at the applied stretch $\lambda\sim1.38$, due to the stress concentration at the edge of the testing sample (see Fig.~\ref{Fig.1}(b)). In contrast to the  case  with a softer matrix, this sample exceeds the stretch of $\lambda\sim1.3$ without exhibiting any necking of the filament. This establishes that the matrix material response influences the onset of necking, and can eliminate the necking instability within the range of performance of the matrix material. It is worth emphasizing that the filament material and the specimen dimensions are the same in both cases \textcolor{black}{and no dynamic effects are observed in our experiments}.

\section{Problem setting and governing equations} \label{sect:problemstatement}
To explain the constitutive sensitivity observed in the previous section, we develop a simplified theoretical framework. We restrict our attention to plane-strain deformation fields and consider an infinitely long film-matrix system. We assume that the film is perfectly adhered to the matrix and we neglect alternative failure modes, such as delamination and cavitation. Both the filament and the matrix are assumed to be incompressible and are represented by a hyperelastic strain energy density. We will show that despite the simplifying assumptions, this model is able to explain the observed behavior and captures also the role of misfit stretch in the film. 


\begin{figure}[H]
    \centering
    \includegraphics[width=1\textwidth]{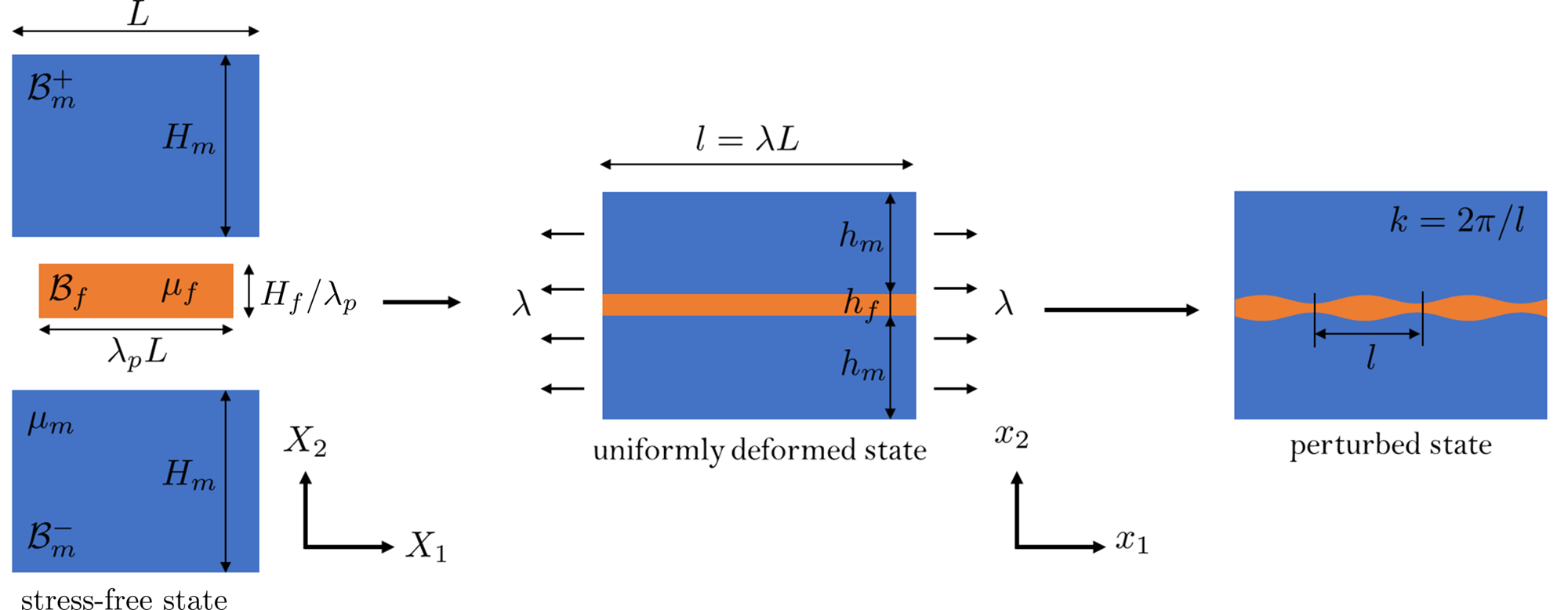}
    \caption{\textbf{Problem settings and coordinate system.} On the left - stress free matrix and stress free film with misfit stretch $\lambda_p$. In the middle - uniformly deformed state. On the right, perturbed state.}
    \label{Fig.2}
\end{figure}

\noindent \textbf{Problem Setting.} Consider a composite system consisting of a thin film of undeformed thickness, $H_f$, embedded in a soft matrix of undeformed thickness, $H_m$; both span indefinitely in the plane of the thin film. We denote the region occupied by the thin film as $\mathcal{B}_f$, and the regions of the matrix on top and bottom sides of the film as $\mathcal{B}^+_m$ and $\mathcal{B}^-_m$, respectively. This multi-layer composite system is deformed by  two different loads (Fig.~\ref{Fig.2}). The first, is an externally applied uniform uniaxial tensile stretch, $\lambda$, that acts along the direction of the film. The second, is an internally activated axial contraction of the film, i.e. a misfit stretch,   $\lambda_p$, which may be induced, for example, by active contraction of cells \citep{wang2013necking, banerjee2015propagating, ban2019strong}, or differential drying \citep{de2008analysis,goehring2013evolving}. The order of application  these two loads does not influence the results in this work.   

Although, both loads subject the film to a tensile stress and it thus is  not expected to exhibit flexural buckling, in this work we ask: \textit{Do periodic solutions exist in this system?} and if so: \textit{At what applied stretch will the system bifurcate and what is the nature and wavelength of this bifurcated response?}

\begin{figure}[ht]
    \centering
    \includegraphics[width=0.7\textwidth]{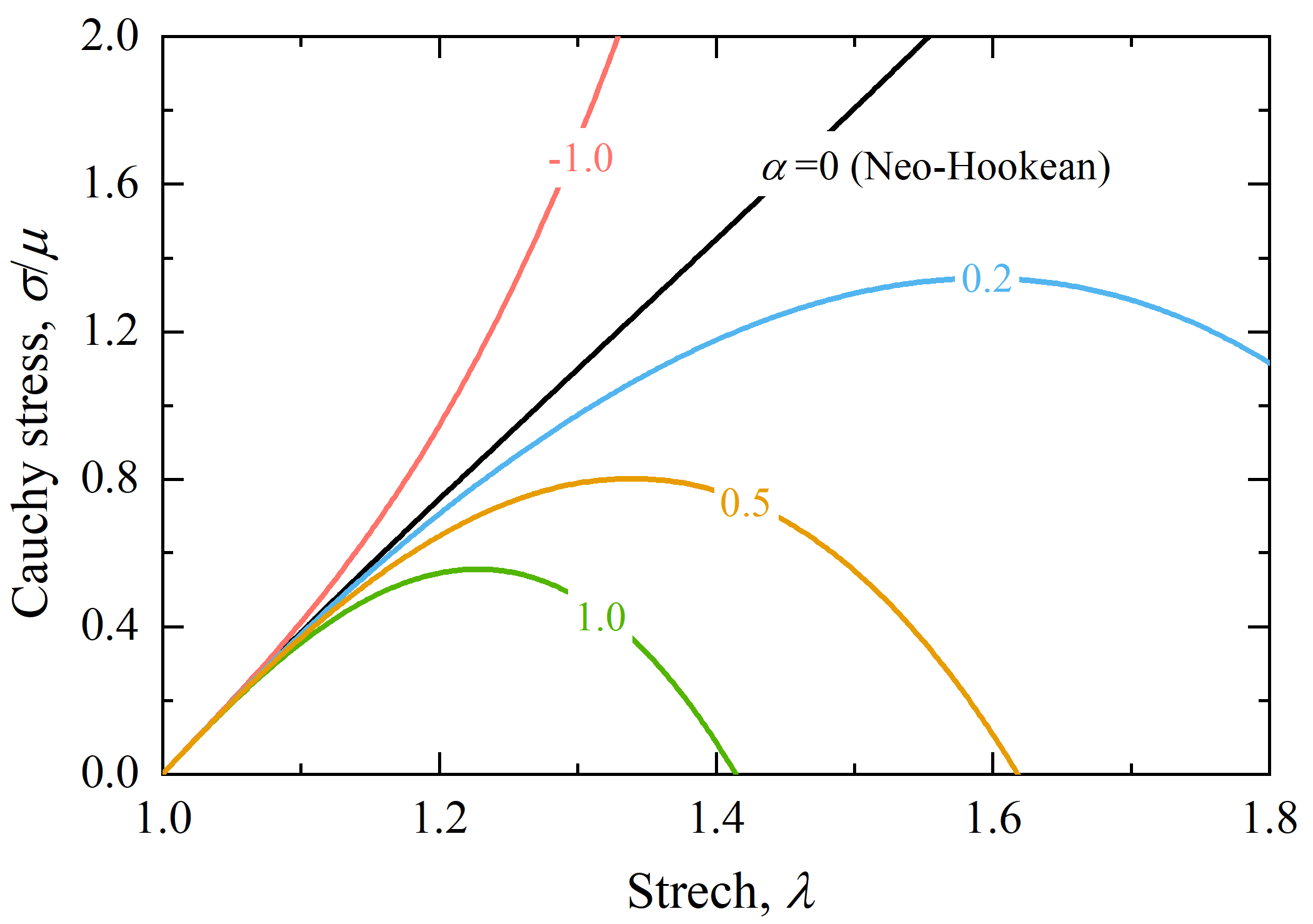}
    \caption{Uniaxial stress-stretch response obtained using the hyperelastic  constitutive model in Eq.  \eqref{psi} for varying values of $\alpha$. }
    \label{Fig.3}
\end{figure}

To answer these questions we assume that both the thin film and the embedding matrix are incompressible and we restrict our attention to plane-strain conditions while neglecting any rate dependent effects. Additionally, we account for the potential onset of tensile failure  of the filament, which can manifest in the form of  softening in the stress-stretch response, as illustrated in Fig.~\ref{Fig.3}.
To capture this behavior, the matrix and the thin film are both modeled using the strain energy density function of the generalized form
\begin{equation}\label{psi}
    \Psi(I_1)=\frac{\mu}{2}\left[I_1-3-\alpha(I_1-3)^2\right],
\end{equation}
where $I_1={\rm tr}({\mathbf{FF}}^{\rm T})$ is the first invariant of the right Cauchy-Green deformation tensor, $\mathbf{F}=\partial {\bf x}/\partial \bf{X}$ is the deformation gradient, and $\bf{x},\bf{X}$ are the Cartesian coordinates of the material point in the current and reference  configurations, respectively. Additionally, the response is described by two material parameters: $\mu$ is the shear modulus in the linear elastic limit, and $\alpha$ is a dimensionless  parameter that defines the material stiffening/softening response (see Fig.~\ref{Fig.3}). This model reduces to the classical neo-Hookean response for $\alpha=0$, while $\alpha<0$, and $\alpha>0$ correspond to strain stiffening and strain softening responses at large deformations, respectively. In this study we will restrict our attention  to  neo-Hookean response in the matrix $(\alpha=0)$, and to softening response in the thin film $(\alpha>0)$. \textcolor{black}{Such  softening  is accompanied by loss of convexity at large deformations and thus implies loss of ellipticity, which is directly associated with the onset of the necking instability in a free-standing film, as will be described in the next section. In this work, we seek to understand the influence of the confining medium on necking and thus examine solutions that bifurcate from the stable branch of the homogeneous deformation response.  Accordingly, we  distinguish the shear moduli of the matrix and the filament, denoted by}  $\mu_m$ and $\mu_f$, respectively; the subscripts $m$ and $f$ will be similarly used throughout the paper to denote quantities in the matrix and the film. Overall, the geometric ratio, $H_f/H_m$, along with the dimensionless constitutive parameters, i.e. the film softening parameter, $\alpha$, and the stiffness ratio, $\mu_f/\mu_m$, provide a complete set of model parameters to  describe the response of the system.

\bigskip 
\noindent \textbf{Uniform pre-deformation.} Considering a system that spans indefinitely along the direction of the thin film  (i.e.  $-\infty<x_1,X_1<\infty$), both the thin film and the matrix are assumed to deform uniformly before onset of the bifurcation. Accordingly, we write the 
ground state deformation gradient and the corresponding first invariant, $I_1$, as
\begin{equation}
\mathbf{F}^0=\left[\begin{matrix}\lambda_1&0&0\\0&\lambda_2&0\\0&0&1\\\end{matrix}\right],\qquad I_1^0(\lambda_1,\lambda_2)=\lambda_1^2+\lambda_2^{2}+1,\end{equation}       where $\lambda_1$ and $\lambda_2$ denote the stretches in the $x_1$ and $x_2$ directions, respectively, and  incompressibility  dictates the relation $\lambda_1\lambda_2=1$. For the matrix, the stretch is determined by the external loading, such that $\lambda_{1}\equiv\lambda$, while in  the thin film the misfit stretch implies $\lambda_1=\lambda\lambda_p$. 

The total potential energy in the uniformly deformed state is thus written as 
\begin{equation}\label{L0}
     \mathcal{L}^0(\lambda_1,\lambda_2)=\int\displaylimits_{\mathcal{B}_m} \left(\Psi_m(I_1^0)-p_m^0(\lambda_1\lambda_2-1)\right){\rm d}V+
     \int\displaylimits_{\mathcal{B}_f} \left(\Psi_f(I_1^0)-p_f^0(\lambda_1\lambda_2-1)\right){\rm d}V
\end{equation}
 where we have used the shorthand notation  $\mathcal{B}_m\equiv\mathcal{B}_m^-\cup \mathcal{B}_m^+$, and we have included the incompressibility constraint in both the matrix and the thin film using the Lagrangian multipliers $p_m^0$ and $p_f^0$, respectively.  
 
 For future use, we derive the longitudinal Cauchy stress in the uniform state, $\sigma_1^0$, from the usual Coleman-Noll methodology on the dissipation rate, which, provided the traction-free condition on the free boundaries of the matrix (i.e. $\sigma_2^0=0$) reads $\mathcal{\dot D}=s_1^0\dot\lambda_1-\mathcal{\dot L}^0(\lambda_1,\lambda_2)=0$, where $s_1^0=\sigma^1_0/\lambda_1$ is the longitudinal stress per undeformed area (i.e. the first Piola stress). Accordingly, by using differentiation by parts, we have  the constitutive relations
 \begin{equation}
     \sigma^0_1=\frac{1}{\lambda_2}\frac{{\rm d}\Psi}{{\rm d}I_1}\left(\frac{\partial I_1}{\partial \lambda_1}\right)+p^0, \qquad p^0=\frac{1}{\lambda_1}\frac{{\rm d}\Psi}{{\rm d}I_1}\left(\frac{\partial I_1}{\partial \lambda_2}\right).
 \end{equation}
 which have the same form for both the matrix and the film. 
 
 We now specialize the above relations by substituting \eqref{psi}, along with the incompressibility constraint $(\lambda_2=1/\lambda_1)$, to write
   \begin{equation}\label{sig0}
     \sigma^0_1=\mu\left[\lambda_1^2-\lambda_1^{-2}-2\alpha\left(\lambda_1^4-\lambda_1^{-4}-2(\lambda_1^2-\lambda_1^{-2})\right)\right],
 \end{equation}
 and 
   \begin{equation}\label{p0}
     p^0={\mu}\lambda_1^{-2}\left[1-2\alpha\left(\lambda_1^{1}-\lambda_1^{-1}\right)^2\right].
 \end{equation}
 Recall that in the above relations, for the matrix we have $\mu=\mu_m$, $\alpha=0$, and $\lambda_1=\lambda$, while for the thin film we have $\mu=\mu_f$,  $\lambda_1=\lambda\lambda_p$, and the softening parameter $\alpha$ can take any positive value. 
 
 \bigskip

\noindent \textbf{Perturbed deformation field.} In this work, to identify the onset of periodic solutions, we consider small perturbations from  the uniform deformation field in the form  
\begin{equation}\label{F}
    {\mathbf{F}}=\left[\begin{matrix}\lambda_1\left(1+u_{1,1}\right)&u_{1,2}/\lambda_1&0\\\lambda_1 u_{2,1}&{\left(1+u_{2,2}\right)}/{\lambda_1}&0\\0&0&1\end{matrix}\right] \quad\text{with}\quad \Delta I_1=\lambda_1^2\left(u^2_{1,1}+u^2_{2,1}+2u_{1,1}\right)+\lambda_1^{-2}\left(u^2_{2,2}+u^2_{1,2}+2u_{2,2}\right).
\end{equation}
Here, without loss of generality and for mathematical convenience, we have prescribed the perturbation  such that it gives rise to the   displacements $U_1(X_1,X_2)=\lambda_1u_1(X_1,X_2)$ and $U_2(X_1,X_2)=u_2(X_1,X_2)/\lambda_1$, along the horizontal and vertical coordinates, respectively, and with $\Delta I_1=I_1-I_1^0$. 
To comply with the incompressibility constraint, we require $\det \mathbf{F}=1$, which from \eqref{F} specialises to the form 
\begin{equation}\label{C}
    C=u_{1,1}+u_{2,2}+u_{1,1}u_{2,2}-u_{1,2}u_{2,1}=0.
\end{equation}

Provided this perturbation, we seek solutions for which the associated energy change is stationary. Hence, we write the corresponding energy change,  $\Delta\mathcal{L}=\mathcal{L}-\mathcal{L}^0$, as
\begin{equation}\label{DL} 
 \Delta \mathcal{L}= \int\displaylimits_{\mathcal{B}_m}\underbrace{ (\Psi(I_{1})-\Psi^0_m-(p_m+p_m^0)C)}_{\mu_m\mathcal{M}(u_{1,1},u_{1,2},u_{2,1},u_{2,2},p_m; \lambda_1)}{\rm d}V+
     \int\displaylimits_{\mathcal{B}_f} \underbrace{(\Psi(I_{1})-\Psi^0_f-(p_f+p_f^0)C)}_{\mu_f\mathcal{F}(u_{1,1},u_{1,2},u_{2,1},u_{2,2},p_f; \lambda_1)}{\rm d}V
\end{equation} where the incompressibility constraint is readily included as in \eqref{L0}, but with the spatially varying Lagrangian multipliers of the perturbed field, $p_m=p_m(X_1,X_2)$ and $p_f=p_f(X_1,X_2)$, for the matrix and the thin film, respectively.  The explicit forms of the integrands, $\mathcal{M}$ and $\mathcal{F}$, can be found in equations \eqref{M_A} and \eqref{F_A} of Appendix B.

Substituting the above functional into the Euler-Largrange equations,  and omitting higher order terms, yields (after some reorganization) the two separate systems of equations, for the matrix region and the thin film region, respectively,
\begin{equation}\label{gov_eq}
    \begin{dcases}
        \lambda_1^2u_{1,11}+\lambda_1^{-2}u_{1,22}-p_{m,1}=0, & \text{in}\quad\mathcal{B}_m\\
        \lambda_1^2u_{2,11}+\lambda_1^{-2}u_{2,22}-p_{m,2}=0,\\
        u_{1,1}+u_{2,2}=0,
    \end{dcases}\quad 
        \begin{dcases}
        (\lambda_1^2\beta+4\alpha(1-\lambda_1^4))u_{1,11}+\lambda_1^{-2}\beta u_{1,22}-p_{f,1}=0, & \text{in}\quad\mathcal{B}_f\\
        \lambda_1^2\beta u_{2,11}+(\lambda_1^{-2}\beta+4\alpha(1-\lambda_1^{-4}))u_{2,22}-p_{f,2}=0,\\
        u_{1,1}+u_{2,2}=0,
    \end{dcases}
\end{equation} where, without loss of generality, we have located the center of the coordinate system at the bottom interface between the filament and the matrix, and  we have substituted $\beta=1-2\alpha(\lambda_1^{1}-\lambda_1^{-1})^2$.  Note that for $\alpha=0$, we have $\beta=1$ and both systems of equations take the same form. 
For completeness, we have included the Euler-Lagrange equations in Appendix B \eqref{E_L},\eqref{A_nBCs}, which includes also  eight natural boundary conditions; four on the free surfaces of the matrix
\begin{equation}\label{nBCs_free}
  \begin{aligned}
        \left.\begin{aligned} u_{1,2}^m+u_{2,1}^m=0\\(4\alpha+\lambda^{-2})u_{1,1}^m+\lambda^{-4}(\alpha-\lambda^{2})u_{2,2}^m+p_m=0\end{aligned} \right\} 
        & \quad\text{on} \quad   
        \begin{aligned} &\partial\mathcal{ B}^+_m-\partial\mathcal{ B}_f \quad (X_2=H_m+H_f)\\ &\partial\mathcal{ B}^-_m-\partial\mathcal{ B}_f \quad (X_2=-H_m)\end{aligned}\\ 
    \end{aligned}
\end{equation}
and four at the interfaces between the regions of the matrix and the thin film
\begin{equation}\label{nBCs_interface}
  \begin{aligned}
          \left.\begin{aligned} 
          \frac{\mu_m}{\mu_f}(u_{1,2}^m+u_{2,1}^m)-\lambda_p^{-2}\beta(u_{1,2}^f+u_{2,1}^f) =0\\
          (4\alpha+(\lambda\lambda_p)^{-2})u_{1,1}^f+(\lambda\lambda_p)^{-4}(\alpha-(\lambda\lambda_p)^2)u_{2,2}^f+p_f \quad~~~~~~ \\
          -\frac{\mu_m}{\mu_f}\left[(4\alpha+\lambda^{-2})u_{1,1}^m+\lambda^{-4} (\alpha-\lambda^2)u_{2,2}^m+p_m\right] =0
          \end{aligned}
          \right\} & \quad\text{on} \quad   
          \begin{aligned} &\partial\mathcal{ B}^+_m\cup\partial\mathcal{ B}_f \quad (X_2=H_f)\\ &\partial\mathcal{ B}^-_m\cup\partial\mathcal{ B}_f \quad (X_2=0)\end{aligned}
    \end{aligned}
\end{equation} 
Here superscripts, `$m$' and `$f$'  denote the displacements in the regions of the matrix and the thin film, respectively. 
 
Finally, displacement continuity at the film--matrix interfaces implies
  \begin{equation}\label{continuity}
    \begin{aligned}
                \left.\begin{aligned} u_{1}^m-u_{1}^f=0\\ u_{2}^m-u_{2}^f=0 \end{aligned}\right\}\quad & \text{on} 
                \quad \begin{aligned} &\partial\mathcal{ B}^+_m\cup\partial\mathcal{ B}_f  \quad (X_2=H_f)\\ 
                &\partial\mathcal{ B}^-_m\cup\partial\mathcal{ B}_f \quad (X_2=0)\end{aligned}\\
    \end{aligned}
\end{equation}

 The governing system of equations \eqref{gov_eq}, along with the boundary and interface conditions \eqref{nBCs_free}-\eqref{continuity} provide a complete set of equations for the three unknown functions $(u_1,u_2,p)$ within the entire domain of the matrix and the thin film. 
 Here we seek  solutions that are periodic along the direction of the film, hence we take
  \begin{equation}\label{generalsolution}
 u_{1}=f(KX_2)\sin(KX_1),\quad u_{2}=g(KX_2)\cos(KX_1), \quad
                 \text{and}\quad  p=Kh(KX_2)\cos(KX_1).
\end{equation}were we have introduced an arbitrary wave number $K$ in the undeformed configuration. 

Substituting the above functions \eqref{generalsolution} into the governing equation \eqref{gov_eq} reads
 \begin{equation}\label{characequ}
                E_4g''''+E_2g''+E_0g=0,
\end{equation}
where $E_0=\lambda_1^2[1-2\alpha(\lambda_1^2+\lambda_1^{-2}-2)]$, $E_2=2\alpha(-3\lambda_1^{4}-3\lambda_1^{-4}+2\lambda_1^{2}+2\lambda_1^{-2}+2)-(\lambda_1^2+\lambda_1^{-2})$, and $E_4=\lambda_1^{-4}E_0$, and the superimposed prime denotes differentiation. The characteristic equation of the differential equation \eqref{characequ} is
 \begin{equation}\label{cccharacequ}
                E_4t^{4}+E_2t^{2}+E_0t=0,
\end{equation}
which has four characteristic roots
 \begin{equation}\label{root}
    t_{1,2}=\pm\sqrt{\frac{-E_2-\sqrt{E_2^2-4E_4E_0}}{2E_4}},
    \quad \text{and} \quad
    t_{3,4}=\pm\sqrt{\frac{-E_2+\sqrt{E_2^2-4E_4E_0}}{2E_4}}.
\end{equation}
Note that for the neo-Hookean limit, i.e. $\alpha=0$, these roots reduce to the well known result  $t_{1,2}=\pm 1$, and $t_{3,4}=\pm\lambda^2$ \citep{biot1963surface,cao2012wrinkles,holland2017instabilities}. 

Provided the characteristic roots \eqref{root}, we can write the general solution of the differential equation \eqref{characequ} in the form
 \begin{equation}\label{gesolu}
  u_2=(Ae^{t_1KX_2}+Be^{t_2KX_2}+Ce^{t_3KX_2}+De^{t_4KX_2})\cos(KX_1),
\end{equation}
where the constants $A, B, C, D$ are to be determined by satisfying the  boundary conditions. Based on the governing equation \eqref{gov_eq}, we can further obtain the general solutions for $u_1$, and $p$. Since the general solutions are applicable to the film, the upper substrate, and the bottom substrate, we have twelve  unknown coefficients. There are four natural boundary conditions on the free surfaces of the matrix, four natural boundary conditions at the film and matrix interfaces. The displacement continuity boundary conditions result in four additional equations on the film and matrix interface. Therefore, substitution of the general solutions into the boundary conditions leads to a set of twelve homogeneous equations for twelve constants with the coefficient matrix \textbf{M} given in Appendix B. To permit nontrivial solutions, the determinant of \textbf{M} must vanish, i.e. $\det\textbf{M}=0$, which determines the bifurcation stretch for a given wave number, $K$.

 \section{Finite element simulation}
\begin{figure}[H]
    \centering
    \includegraphics[width=0.7\textwidth]{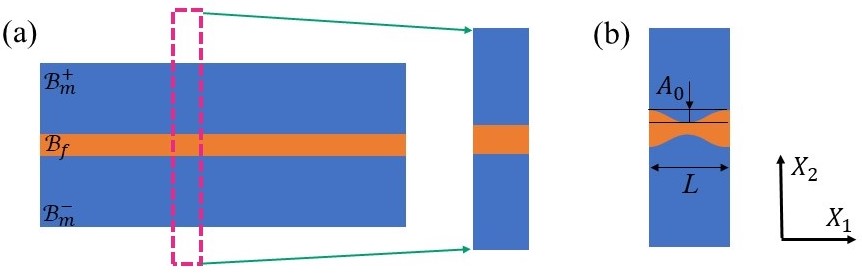}
    \caption{\textbf{Numerical setting.} (a) Construction of unit cell and, (b) illustration of geometrical imperfections.}
    \label{Fig.4}
\end{figure}

 To numerically predict the onset of film necking in a matrix, and to obtain the corresponding critical stretch and critical  wavelength, we employ the Bloch wave instability analysis, which is implemented in the \textcolor{black}{explicit nonlinear} finite element code COMSOL. \textcolor{black}{Each mesh is constructed by cubic, square plane strain elements. A mesh sensitivity analysis is performed to ensure  accuracy of the results.} For numerical convenience, we employ a nearly incompressible material model, and thus the strain energy density of the film (Eq. \ref{psi}) is  modified to include an energy penalty for any deformation that is not isochoric. Plane-strain conditions are assumed as in the previous section. In our simulation, we first apply the macroscopic deformation of the sample (i.e. $\lambda$) to the unit cell (Fig.~\ref{Fig.4}) by imposing periodic displacement boundary conditions, with the  contraction of the film modeled as a pre-stretch. Next, we apply the Bloch-Floquet boundary conditions on the deformed unit cell. We solve the corresponding eigenvalue problem for a given wavenumber $K$ at applied deformation $\lambda$. Once a non-trivial zero eigenvalue is detected, the corresponding wavenumber and applied deformation are identified as the critical wavenumber - $K_{cr}$ and critical stretch - $\lambda_{cr}$. For more details on the Bloch-Floquet instability analysis we refer the reader to earlier studies \citep{bertoldi2008wave,slesarenko2017microscopic}. 
 
Finally, we perform a post-bifurcation analysis  to investigate the film necking process. A unit cell of length $L$ is constructed. Note that $L=2\pi /K$ is the bifurcated wavelength in the undeformed state, $K$ is the undeformed wavenumber. Compressive deformation is applied by imposing periodic displacement boundary conditions on the constructed unit cell, along the $X_1$ direction. Small amplitude geometrical imperfections in the form of $X_2=A_0 \cos{(2\pi X_1/L)}$ are imposed on the film-matrix interface. In particular, an amplitude of $A_0 / H_f=5\times10^{-4}$ is  found to be sufficient to trigger the necking bifurcation.
 
  \section{Results and discussion}
  In this section, we  present  analytical results for the tensile necking of a softening film embedded in a matrix, and compare the  analytical analysis  with finite element simulation results obtained by the  Bloch wave instability analysis and the post-buckling analysis. We then discuss the effect of model parameters on necking behavior of the composite system.
  
\begin{figure}
    \centering
    \includegraphics[width=0.9\textwidth]{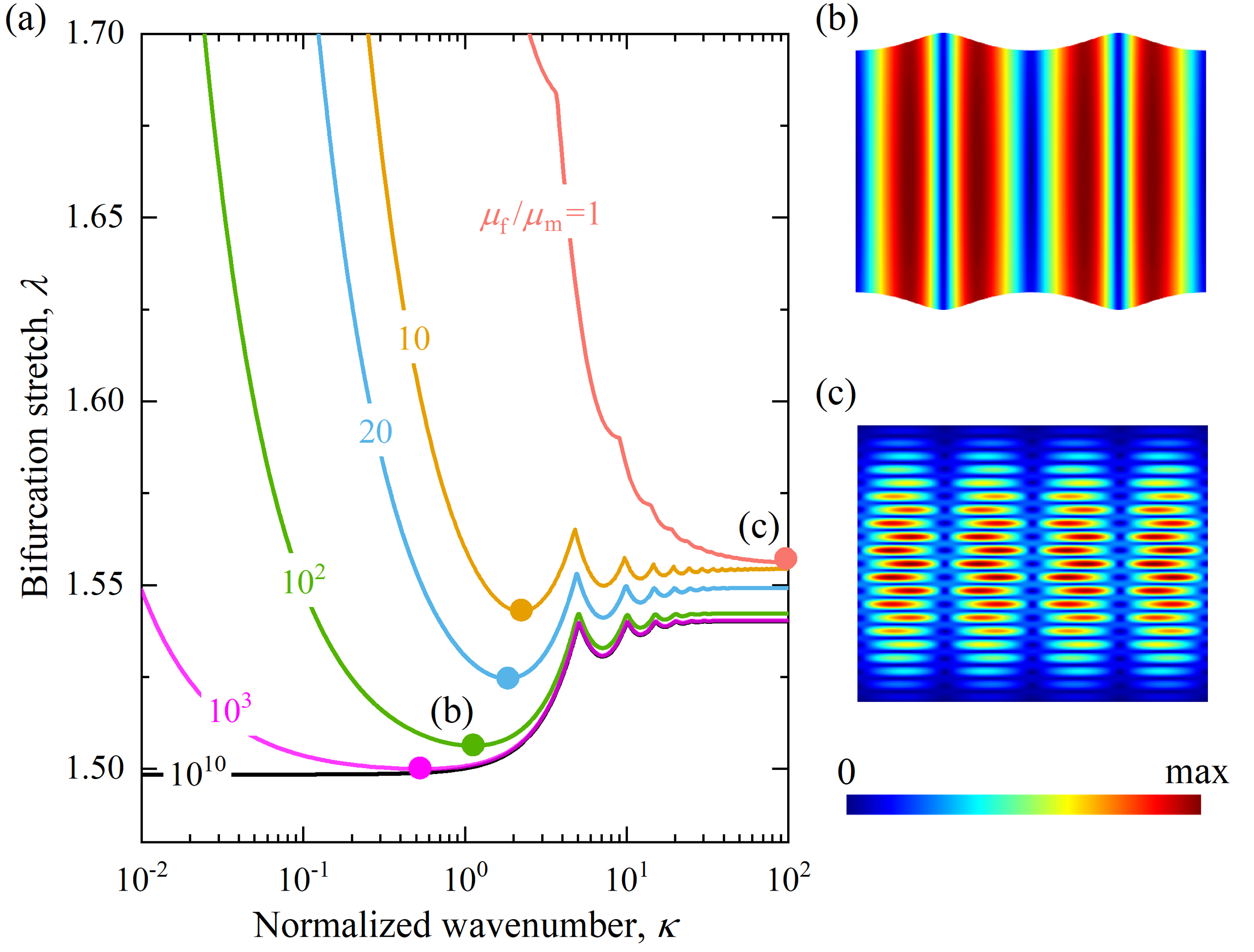}
    \caption{\textbf{Analytically  predicted bifurcation stretch and bifurcation modes.} (a) Bifurcation stretch verse normalized wavenumber for the composite system with different shear modulus contrasts. Corresponding bifurcation modes are shown in (b) for the shear modulus contrast $\mu_f/\mu_m =1$, and in (c) for $\mu_f/\mu_m =100$. Here we consider infinite matrix thickness and no misfit stretch  with  the film softening $\alpha=0.2$. The filled circular points represent the lowest buckling stretch, defined as the critical stretch in this work. The corresponding normalized wavenumber is defined as the critical wavenumber. The color in (b) and (c) represents the total displacement of the bifurcation mode. Note that $\kappa=KH_f$, where $K$ is the undeformed wave number. }
    \label{Fig.5}
\end{figure}

  We first present theoretical predictions of the critical bifurcation stretch and its dependence  on the  normalized wavenumber. As an example, we consider the composite system with infinite matrix thickness $(H_m/H_f\to\infty)$ and no misfit stretch. Representative results obtained for the film softening coefficient of $\alpha=0.2$, are shown in Fig.~\ref{Fig.5}, for a wide range of shear modulus contrasts. We find that the critical bifurcation stretch, defined as the global minimum for each stiffness contrast (as marked by filled circular points) is observed at a finite wavenumber for moderate shear modulus contrasts, i.e. $\mu_f/\mu_m=10, 20, 10^2,10^3$. Moreover, the higher the contrast, the lower critical bifurcation stretch and  the smaller the corresponding critical wavenumber. A typical bifurcation mode is shown in Fig.~\ref{Fig.5}(b). Consistent with the necking instability, we observe that the bifurcation mode is symmetric in the film thickness direction, and that the displacement along the film thickness is highly uniform. For the composite system with higher shear modulus contrast, we observe that the critical stretch emerges at plateau  region of the  curve  (see the case with $\mu_f/\mu_m=10^3$), implying that the buckling wavelength for the composite with stiffer film would be more sensitive to imperfections, similar to the observations of finite wavelength domain patterns in periodic laminate composites \citep{li2022emergence}, and the herringbone pattern in  bilayer systems \citep{chen2004family,chen2004herringbone}. 
  
  For the free standing softening film, i.e. $\mu_f/\mu_m\rightarrow\infty$, following the Considère criterion \citep{Considerecrete1885}, the necking stretch can be calculated by
  \begin{equation}\label{consider}
  d\sigma/\sigma=-dA/A,
\end{equation}
where $\sigma$ is the Cauchy stress in uniaxial tension and $A$ is the instantaneous cross-sectional area.  From material incompressibility we can write
\begin{equation}\label{considercre}
  dA/A=-d\lambda/\lambda,
\end{equation}
Combining Eqs. \eqref{consider} and \eqref{considercre} and substituting relation \eqref{sig0},  we write  a criteria for the necking stretch for the free-standing softening film as 
\begin{equation}
    \lambda^2+3\lambda^{-2}-2\alpha\left(3\lambda^4+5\lambda^{-4}-2(\lambda^2+3\lambda^{-2})\right)=0.
\end{equation} 
Accordingly, for the case with $\alpha=0.2$, the necking stretch  $\lambda_{cr}=1.4984$ is obtained numerically. It is worth mentioning that according to this criteria, there is a direct correlation between material softening and necking (i.e. necking would occur if $\alpha>0$).As shown in Fig.~\ref{Fig.5}(a), when $\mu_f/\mu_m$  is large, the effect of the compliant matrix becomes negligible and the critical bifurcation stretch asymptotically approaches  the necking stretch of the free standing film.

For  soft films, the critical bifurcation stretch occurs at the limit of short wavelengths, i.e. $\kappa=KH_f\rightarrow\infty$. A typical bifurcation mode in Fig.~\ref{Fig.5}(c) shows that the maximum displacement occurs at the center cord of the film and decreases along the film thickness direction, indicating the failure might nucleate from defects in the bulk of the film in the form of cavitation,  reminiscent to the classical work of \cite{ashby1989flow}. In contrast, for the case with a stiffer film,  the damage is expected to first occur at the film-matrix interface. Moreover, we find that in the short wave limit, the bifurcation stretch increases with decreasing shear modulus contrast.

Comparing the analytical predictions  with experimental observations in Fig.~\ref{Fig.1}, reveals a qualitative agreement. Both experimental and analytical results show that the composite with a higher film-to-matrix shear modulus contrast bifurcates earlier. For moderate contrast, the film exhibits necking at a finite wavelength. For the softer film, although no macroscopic cracks are observed before the failure of the specimen in experiments, micro-cracks may accumulate internally to accommodate the large strains, as predicted by the theoretical analyses.

\begin{figure}[H]
    \centering
    \includegraphics[width=1\textwidth]{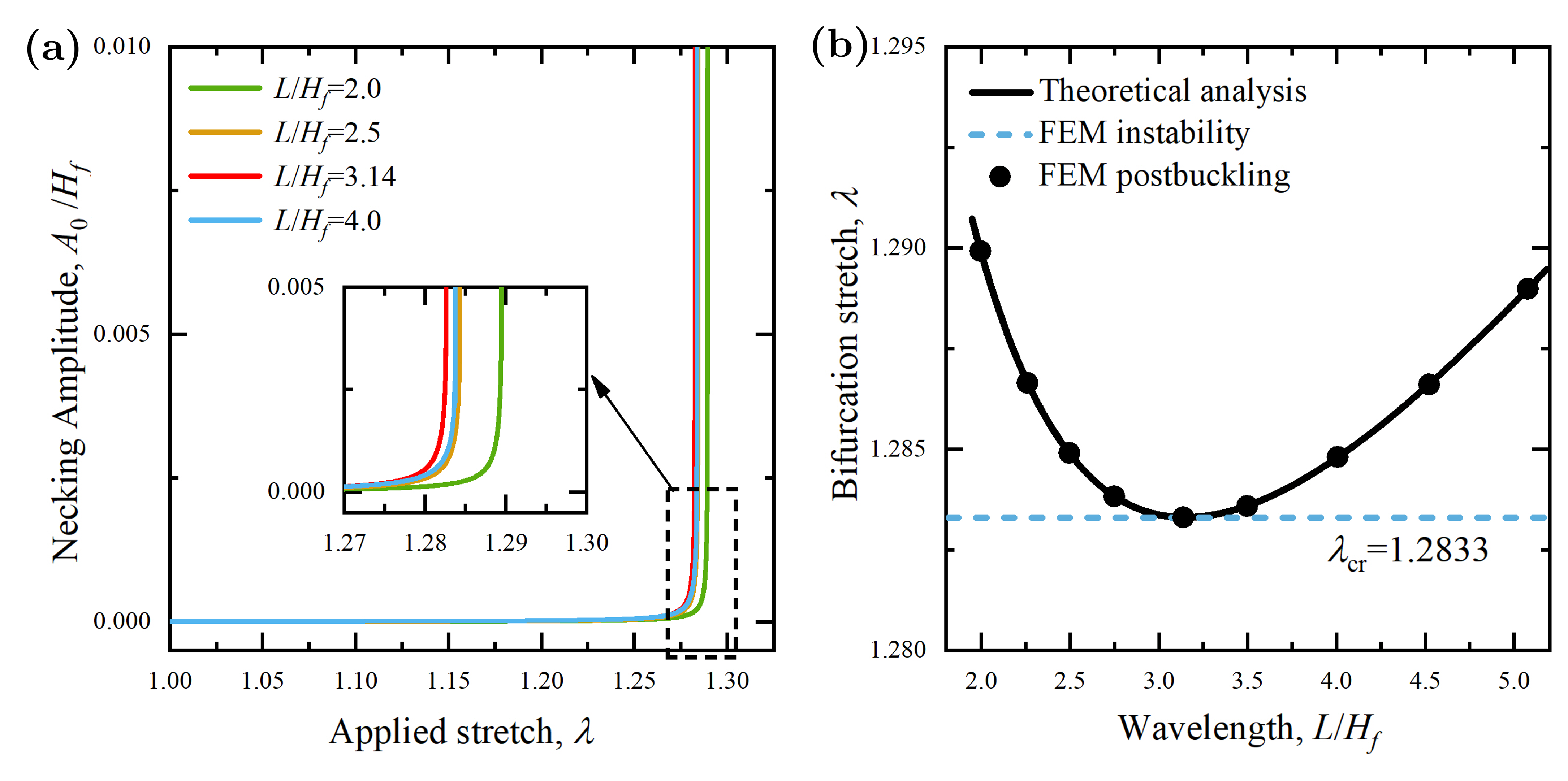}
    \caption{ \textbf{Comparison of theoretical predictions and finite element simulation results.} (a) Dependence of necking amplitude on applied deformation obtained by finite element post-bifurcation analysis with various wavelengths $L/H_f$. (b) Dependence of necking stretch on bifurcation wavelength. Results are given for  composite with $H_m/H_f\to\infty$, $\alpha=0.2$, $\mu_f/\mu_m =10$, and $\lambda_p=1.2$. The critical stretch $\lambda_{cr}$ is the one for which the stretch is minimal, the corresponding wave length $L_{cr}$ is identified as the critical wavelength, and the critical wavenumber is $K_{cr}=2\pi /L_{cr}$ .}
    \label{Fig.6}
\end{figure}  

To examine the role of necking in the post-bifurcation process we show in Fig.~\ref{Fig.6}(a)  the normalized necking amplitude, $A_0/H_f$, as a function of applied stretch, obtained from  finite element simulations that are conducted for different normalized wavelengths, $L/H_f$. 
Here, we consider a material with $\mu_f/\mu_m =10$ and with $\alpha=0.2$. In contrast to the case in Fig.~\ref{Fig.5}, for which $\lambda_{cr}=1.5428$ and $L_{cr}/H_f=2.84$, we now introduce a film pre-stretch of $\lambda_p$=1.2.
We observe that upon approaching a critical level of applied stretch  the necking amplitude increases sharply, indicating the onset of necking. Moreover, this critical stretch significantly depends on the wavelength. To further examine this wavelength dependence,  Fig.~\ref{Fig.6}(b) presents the corresponding critical stretch as a function of $L/H_f$, and compares with  both the analytical  predictions and the finite element instability analysis. We find that the bifurcation stretch predicted by the theoretical analysis is in excellent agreement with the results of the post-bifurcation analysis. The critical stretch of both (i.e. the minimum stretch, $\lambda_{cr}=1.2833$) agrees well with the value obtained from the numerical instability analysis. In comparison with the critical stretch in absence of pre-stretch (i.e. $\lambda_p=1$) we see an earlier transition to necking, as expected, but with a larger critical wave length $L_{cr}=3.14$. 
Next, we expand the investigation of the role of the model parameters on the composite  stability limit. Having established the agreement between the analytical and the numerical modeling approaches, the following results will be derived using the analytical framework. 

\begin{figure}[H]
    \centering
    \includegraphics[width=1\textwidth]{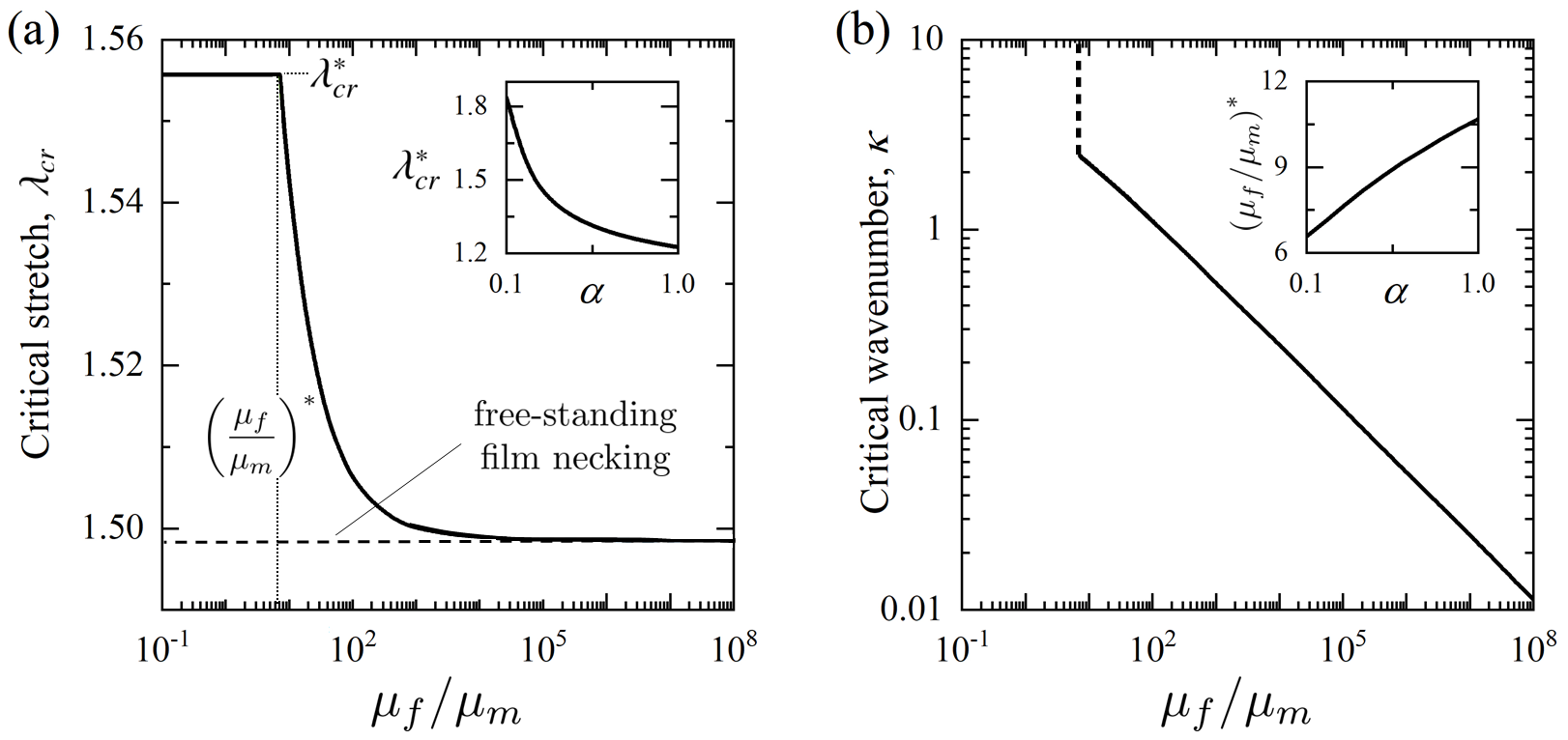}
    \caption{\textbf{Effect of shear modulus contrast and softening coefficient.} Dependence of (a) the critical stretch, and (b) the critical wavenumber,  on the film-to-matrix shear modulus contrast for the composite with $H_m/H_f\to\infty$, $\lambda_p=1$, and $\alpha=0.2$. The transition between the short wavelength bifurcation mode and necking is indicated by $(\mu_f/\mu_m)^*$, with corresponding critical transition stretch $\lambda_{cr}^*$. The dependence of the transition point on the softening parameter $\alpha$ is shown in the insets. From these results it is apparent that the softening response can directly influence not only the critical bifurcation stretch but also the observed bifurcation mode.  }
    \label{Fig.7}
\end{figure} 

\noindent\textbf{Constitutive sensitivity.} The roles of the film-to-matrix shear modulus contrast $(\mu_f/\mu_m)$ and the softening coefficient $\alpha$ in determining the necking response are considered next.   Fig.~\ref{Fig.7} presents the critical stretch and the critical wavenumber as functions of shear modulus contrast for the composite with $H_m/H_f\to\infty$, no misfit stretch ($\lambda_p=1$), and $\alpha=0.2$. We observe that while the dependence of the critical stretch on the  modulus contrast is highly nonlinear, the dependence of the  critical wavenumber on the modulus contrast can be well capture by a power law (as seen from the linear dependence in the log-log plot). As expected, for  $\mu_f/\mu_m\rightarrow\infty$ the critical stretch asymptotically approaches the Considère necking stretch for the free-standing film ($\lambda_{cr}=1.4984$)   with $L_{cr}/H_f\to\infty$. From this curve we can see clearly the transition to the short wavelength bifurcation mode  for small stiffness contrasts  $\mu_f/\mu_m<(\mu_f/\mu_m)^*$. Within this range  necking will not be observed, the critical bifurcation stretch is independent of the modulus contrast, and $\kappa\rightarrow\infty$. \textcolor{black}{Interestingly, in the range of $\mu_f/\mu_m\gg1$, the critical wavenumber remains highly dependent on the fiber-to-matrix stiffness contrast, while the critical stretch asymptotically approaches the Considère limit for the free-standing film. This  implies that while the  wavelength is primarily determined by the stiffness ratio, onset of bifurcation can be highly sensitive to initial imperfections, such as material inhomogeneity or geometrical irregularities.  This sensitivity can explain the observation of sequential necking and fragmentation in Section 2, and in other material systems, such as glass fiber reinforced composites in \citep{shabahang2016controlled} and metal reinforced composites in \citep{cooper2019toughening}. }

The dependence of the  critical modulus contrast $(\mu_f/\mu_m)^*$, and the corresponding critical transition stretch $\lambda_{cr}^*$ on the softening coefficient are shown in the subfigures of Fig.~\ref{Fig.7}. It is observed that an increase in film softening coefficient results in a decrease in the critical transition stretch, and an increase of the critical modulus contrast. \textcolor{black}{We would like to emphasize that while previous studies have  examined the sensitivity of  necking to the stiffness contrast, as for example in the recent work by \cite{jia2019bifurcation}, such studies consider metal sheets that are modeled using $J_2$ deformation theory of plasticity, and thus exhibit distinct necking behaviour compared to hyperelastic materials, as explained by \cite{audoly2016analysis}. Here, by using a general hyperelastic constitutive relation, we obtain a comprehensive understanding of this sensitivity across a broad range of model parameters.}

\begin{figure}[H]
    \centering
    \includegraphics[width=1\textwidth]{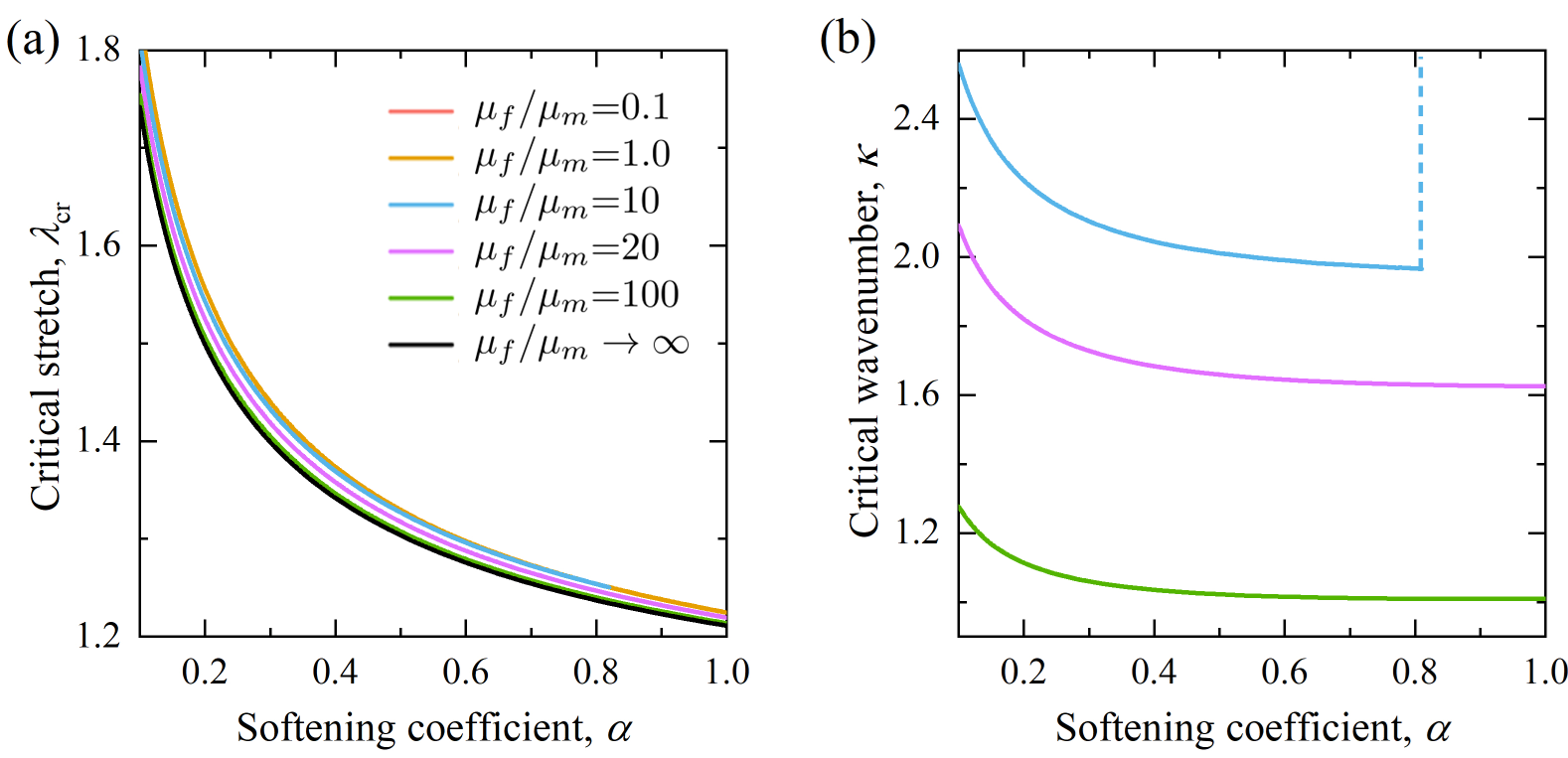}
    \caption{\textbf{Effect of film softening coefficient on critical stretch.} (a) and critical wavenumber (b) for the composite with $H_m/H_f\to\infty$ and $\lambda_p=1$.}
    \label{Fig.8}
\end{figure} 

To further examine  the sensitivity of film bifurcation to the softening coefficient, 
   Fig.~\ref{Fig.8} shows the influence of  $\alpha$ on the critical stretch and wavenumber.  Interestingly, the change of $\alpha$ could trigger a transition of bifurcation mode, from a finite wavelength to short wavelength. For example, the composite with $\mu_f/\mu_m=10$ transitions at $\alpha=0.81$. In addition, it should be noted that the effect of the softening coefficient on critical stretch is much more pronounced than that of the modulus contrast. 
 
 \bigskip

\begin{figure}[H]
    \centering
    \includegraphics[width=1\textwidth]{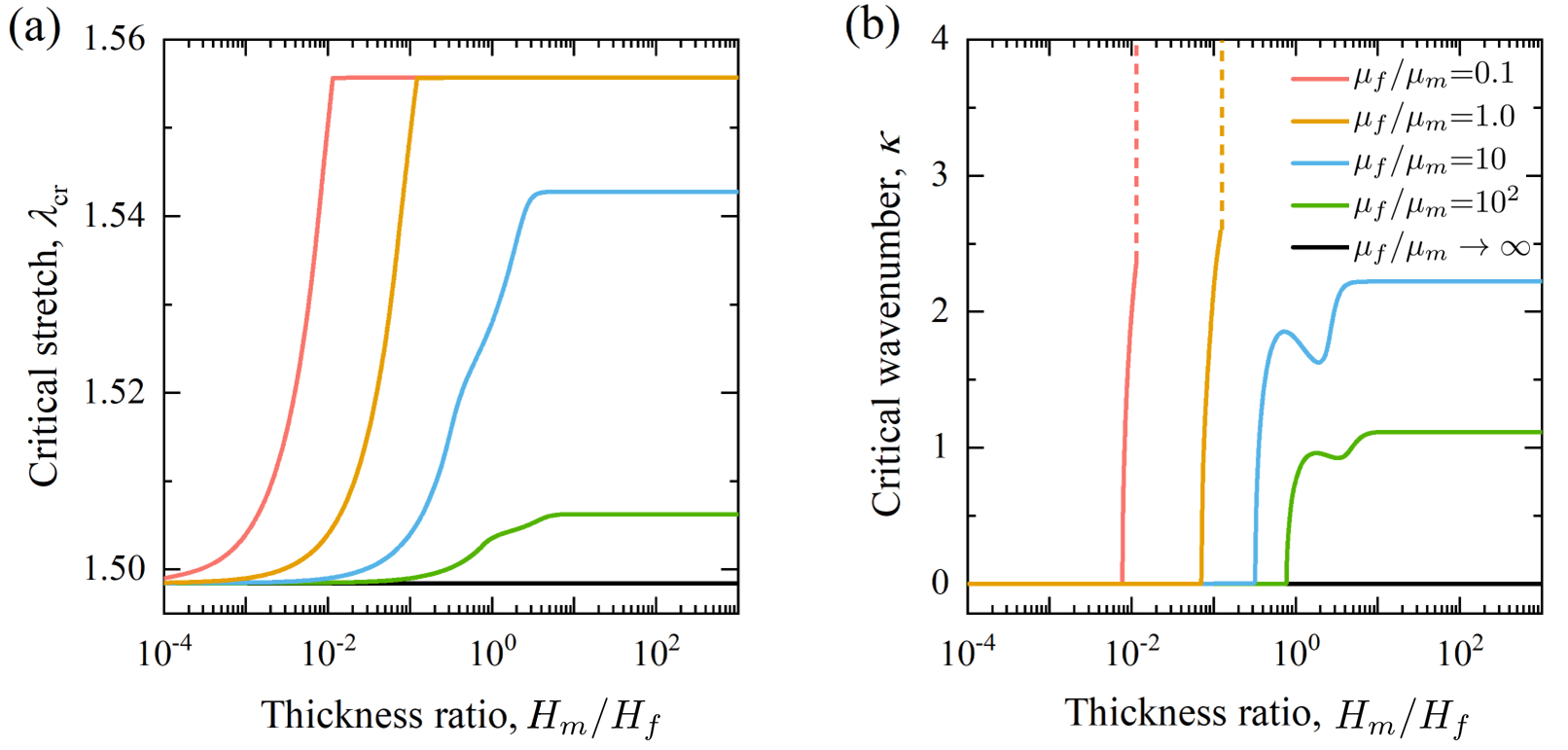}
    \caption{\textbf{Effect of finite matrix thickness.} Dependence of (a) critical stretch and (b) critical wavenumber, on matrix-to-film thickness ratio for the composite with $\lambda_p=1$ and $\alpha=0.2$.}
    \label{Fig.9}
\end{figure}

\noindent\textbf{Geometric Sensitivity.}
Fig.~\ref{Fig.9} shows the critical stretch and critical wavenumber as functions of the matrix-to-film thickness ratio, $H_m/H_f$, for  composite systems with no  misfit stretch, and with $\alpha=0.2$.  We first observe that at the limits $H_m/H_f\rightarrow 0$ and $H_m/H_f\rightarrow\infty$, the critical stretch and the critical wavenumber become independent of thickness ratio. At the former limit, the matrix becomes negligible and we retrieve response of a free-standing film. At the latter limit, the edge of the matrix exceeds the region of influence of the film. 
We also observe that the matrix stabilizes the film, as seen from the   monotonic increase of the critical stretch with  increasing in matrix thickness. However, the dependence of the wavenumber on matrix thickness exhibits non-monotonic behavior.  While the overall trend is that a  thicker matrix bifurcates  at larger wavenumbers, local minima are observed, as for example  at $H_m/H_f =1.9$ for the case with $\mu_f/\mu_m=10$. We also observe that the thickness ratio influences the transition point from short wavelength response to necking  response, as see for $\mu_f/\mu_m=0.1$, and $1$.

\begin{figure}[H]
    \centering
    \includegraphics[width=1\textwidth]{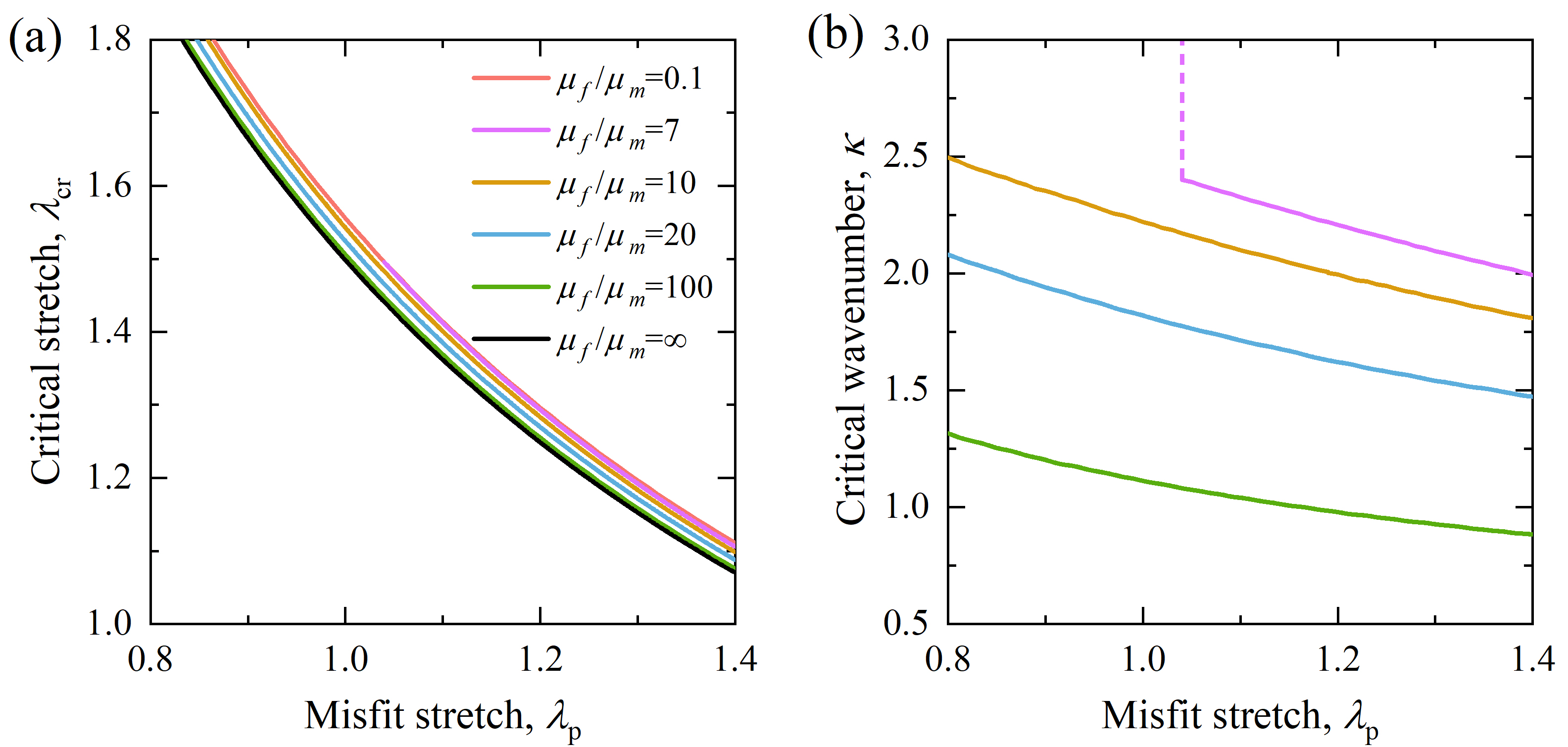}
    \caption{Effect of film  misfit stretch on (a) the critical the stretch and (b) the critical wavenumber. Results are shown  for the composite with $H_m/H_f\to\infty$ and $\alpha$=0.2.}
    \label{Fig.10}
\end{figure}

  \noindent\textbf{Misfit stretch.}
  Finally, we examine the effect of film  pre-deformation on the composite bifurcation behaviors. Such  misfit stretch, has been shown to emerge in biological systems due to active cell contraction, and can appear in synthetic systems as a residual stress that emerge in the fabrication process of multi-material systems. In Fig.~\ref{Fig.10} we show  the critical stretch and critical wavenumber as functions of applied pre-stretch, $\lambda_p$. We find that both the critical stretch and the critical wavenumber  decrease with increasing misfit stretch. The misfit stretch significantly affects the critical stretch. This can be intuitively  explained by the fact that the pre-stretch brings the system closer to the necking limit, thus requiring less stretch for bifurcation. Less intuitive is the influence of pre-stretch on the critical wavenumber. Moreover, we observe that the film misfit stretch can trigger the transition from a short wavelength to a necking bifurcation with finite wavelength, see for $\mu_f/\mu_m=7$.

\section{Summary and concluding remarks}
A hyperelastic filament (or thin film) embedded in a hyperelastic matrix can exhibit bifurcation behaviour in tension that is distinct from the reported behavior in metal films. Our  bifurcation analysis accounts for  strain softening of the filament and for filament pre-stress and exposes three bifurcation modes:
\textit{(i)} A single necking mode for which the deformation is localized to one neck. This mode is  observed in an isolated film or for a weak matrix (i.e. a thin matrix and/or high film-to-matrix stiffness ratio), thus recovering the Considère criterion for necking. \textit{(ii)} A periodic necking mode,  with finite wavelength is observed in the composite with moderate to high film-to-matrix stiffness ratios. This bifurcation mode (i.e. the critical stretch and corresponding wavelength) is shown to be highly sensitive to the stiffness contrast and the strain-softening coefficient.      \textit{(iii)} An internal  softening mode is  observed for small film-to-matrix stiffness ratios. This short wavelength mode is characterized by localized softening at the center cord of the film. Consistent with earlier studies for metal films, we find that the confining matrix delays the onset of  bifurcation; the stiffer the film and the thinner the matrix thickness, the lower the necking stretch. In particular, the internal softening mode is  found to be independent of the film-to-matrix stiffness ratio.  We observe that the softening coefficient has the most pronounced effect on the necking stretch, while the increases of film misfit stretch leads to periodic necking  with longer wavelengths. It is worth  pointing out that although the current work focuses on  softening embedded films that are subjected uniaxial tensile deformation in plane-strain conditions, the formulated theory can be extended to general nonlinear constitutive laws and more general deformations such as uniaxial tension and arbitrary biaxial tensile loading.

 This work is not without limitations. The presented  theory is limited to predict the onset of bifurcation for the film in a hyperelastic matrix with perfect bonding. The transition from necking to rupture \citep{fu2021necking}, the roles of  \textcolor{black}{interfacial delamination \citep{jia2021effect,kothari2020controlled,HenzelPNASn,ringoot2021stick}, interfacial  mixing \citep{gao2017tuning,liu2020effect}}, material rate dependence \citep{chockalingam2021probing}, and dynamic loading \citep{audoly2019one,rodriguez2017critical} require further investigation. Additionally,  more detailed experiments are required to quantify the nonlinear stress softening coefficient and guide the development of more advanced constitute laws for a given material system. It should be noted that the role of imperfections on the bifurcation and post-bifurcation behavior is an interesting direction to explore further. The weak dependence of the critical  bifurcation stretch on the  wavelength, for  large film-to-matrix stiffness ratios, suggests that onset of periodic necking in stiff films can be significantly affected by geometrical imperfections.

\section*{Acknowledgements}
The authors acknowledge the support of Dr. Timothy B. Bentley, Office of Naval Research Program Manager, under award number N00014-20-1-2561. H.V. acknowledges the support of the Department of Defense (DOD) through the National Defense Science \& Engineering Graduate (NDSEG) Fellowship Program. 
\section*{Appendix A}
\setcounter{figure}{0}
\renewcommand{\thefigure}{A\arabic{figure}}
Experimental measurement of stress-stretch response and fitting to obtain the corresponding shear moduli. 

\begin{figure}[H]
    \centering
    \includegraphics[width=1\textwidth]{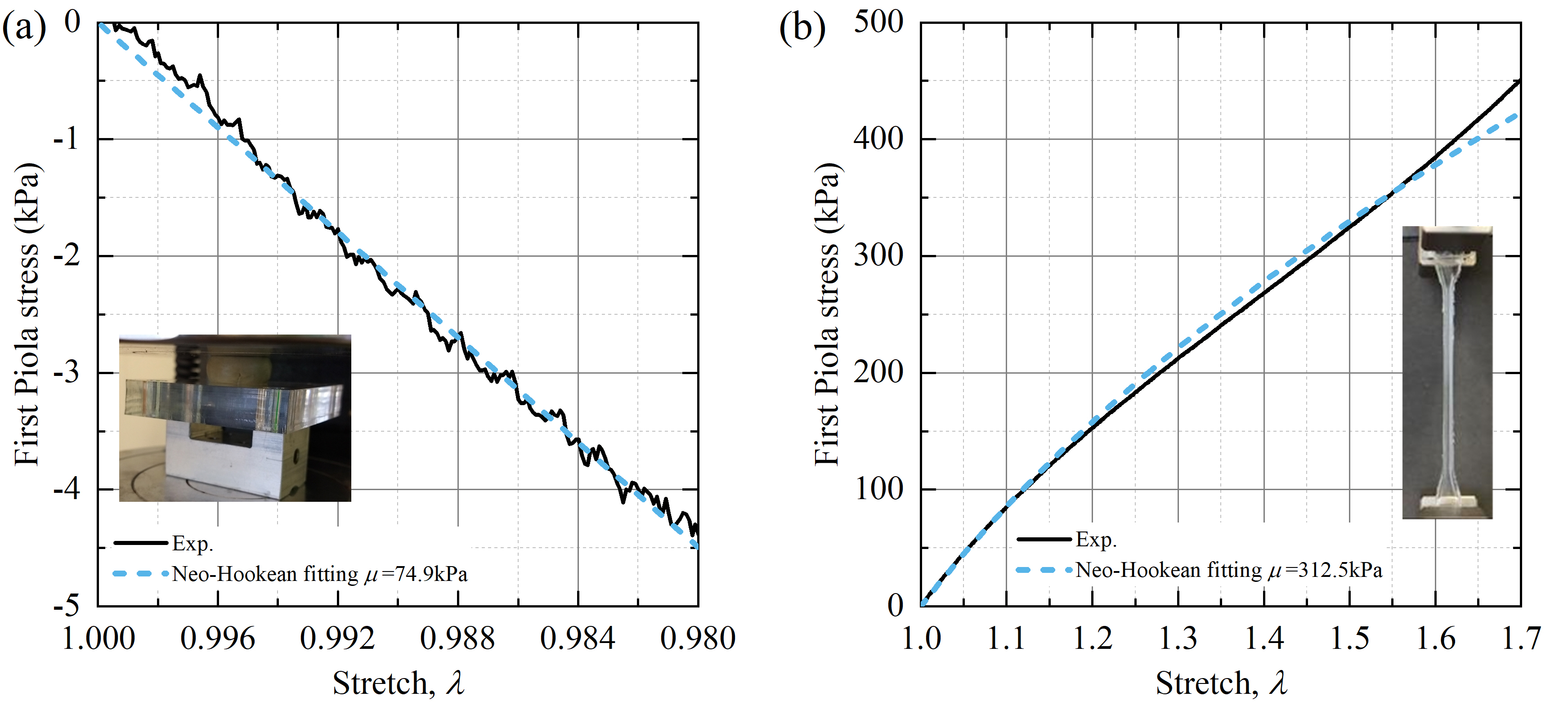}
    \caption{Stress-stretch behavior of (a) putty, and (b) PDMS with base:cross-linker mass ratio 15:1. All experiments are performed at a stretch rate of $1\times10^{-3} s^{-1}$. The shear moduli are obtained through fitting the experimentally obtained stress-stretch curve to an incompressible neo-Hookean model.
    }
    \label{Fig.A1}
\end{figure}

\section*{Appendix B}
\renewcommand{\thefigure}{B\arabic{figure}}
Experimental observation of a single free-standing putty filament of circular cross-section subjected to a tensile deformation along its axis.

\begin{figure}[H]
    \centering
    \includegraphics[width=1\textwidth]{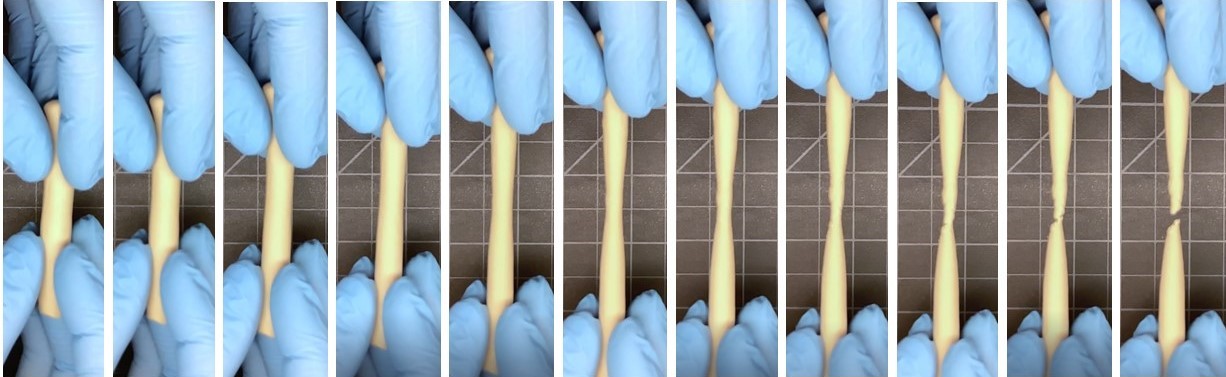}
    \caption{Deformation sequences of a single  putty filament of circular cross-section subjected to tensile deformation. The filament is stretched manually along its axial direction. The development of necking instability is clearly observed and implies that material softening occurs, according to the Considère criterion. This process is repeatable. Nonetheless, the neck can form at any arbitrary location and is highly sensitive to initial imperfections. Only one representative sequence is shown. }
    \label{Fig.B1}
\end{figure}

\section*{Appendix C}
\newcounter{defcounter}
\setcounter{defcounter}{0}
\newenvironment{myequation}{%
        \addtocounter{equation}{-1}
        \refstepcounter{defcounter}
        \renewcommand\theequation{B\thedefcounter}
        \begin{equation}}
{\end{equation}}

\setcounter{figure}{0}
\renewcommand{\thefigure}{B\arabic{figure}}

The integrands in the functional \eqref{DL} are derived by inserting \eqref{psi},  \eqref{p0}-\eqref{C}, which upon removing higher order terms and along with the constraint $\lambda_2=1/\lambda_1$ reads
\begin{myequation}\label{M_A}\begin{split}
\mathcal{M}(u_{1,1},u_{1,2}&,u_{2,1},u_{2,2},p_m; \lambda_1)=\Delta I_1/2-(1/{\mu_m})(p_m+p_m^0)C\\
=&\frac{1}{2}\biggl( \lambda_1^2\left(u^2_{1,1}+u^2_{2,1}+2u_{1,1}\right)+\lambda_1^{-2}\left(u^2_{2,2}+u^2_{1,2}+2u_{2,2}\right)\biggr)\\
&-\left(\frac{p_m}{\mu_m}+\frac{1}{\lambda_1^{2}}\right)\left(u_{1,1}+u_{2,2}+u_{1,1}u_{2,2}-u_{1,2}u_{2,1}\right)
\end{split}
\end{myequation}

\noindent  and 
\begin{myequation}\label{F_A}\begin{split}
\mathcal{F}&(u_{1,1},u_{1,2},u_{2,1},u_{2,2},p_f; \lambda_1)=(\Delta I_1/2)(1-\alpha(\Delta I_1+2I_0-6))-(1/\mu_f)(p_f+p_f^0)C \\
=& \frac{1}{2}\biggl( \lambda_1^2\left(u^2_{1,1}+u^2_{2,1}+2u_{1,1}\right)+\lambda_1^{-2}\left(u^2_{2,2}+u^2_{1,2}+2u_{2,2}\right)\biggr)\Biggl(1+2\alpha\biggl(2- \lambda_1^2\left(u_{1,1}+1\right)
-\lambda_1^{-2}\left(u_{2,2}+1\right)\biggr)\Biggr)\\&-\Biggl(\frac{p_m}{\mu_f}+\lambda_1^{-2}\left(1-2\alpha\left(\lambda_1-\lambda_1^{-1}\right)^2\right)\Biggr)\left(u_{1,1}+u_{2,2}+u_{1,1}u_{2,2}-u_{1,2}u_{2,1}\right)\\
\end{split}\end{myequation}

A stationary solution for the potential energy change in \eqref{DL} is readily obtained by method of variations to arrive at the generic form of the Euler-Lagrange equations, which is provided here for completeness
\begin{myequation}\label{E_L}
    \begin{dcases}
        \frac{{\rm d}}{{\rm d}X_1}\left(\frac{\partial \mathcal{M}}{\partial u_{1,1}}\right)+  \frac{{\rm d}}{{\rm d}X_2}\left(\frac{\partial \mathcal{M}}{\partial u_{1,2}}\right)=0, & \text{in}\quad\mathcal{B}_m\\
        \frac{{\rm d}}{{\rm d}X_1}\left(\frac{\partial \mathcal{M}}{\partial u_{2,1}}\right)+  \frac{{\rm d}}{{\rm d}X_2}\left(\frac{\partial \mathcal{M}}{\partial u_{2,2}}\right)=0,\\
        \frac{\partial \mathcal M}{\partial p_m}=0,
    \end{dcases}\qquad 
        \begin{dcases}
        \frac{{\rm d}}{{\rm d}X_1}\left(\frac{\partial \mathcal{F}}{\partial u_{1,1}}\right)+  \frac{{\rm d}}{{\rm d}X_2}\left(\frac{\partial \mathcal{F}}{\partial u_{1,2}}\right)=0, & \text{in}\quad\mathcal{B}_f\\
        \frac{{\rm d}}{{\rm d}X_1}\left(\frac{\partial \mathcal{F}}{\partial u_{2,1}}\right)+  \frac{{\rm d}}{{\rm d}X_2}\left(\frac{\partial \mathcal{F}}{\partial u_{2,2}}\right)=0,\\
        \frac{\partial \mathcal F}{\partial p_f}=0,
    \end{dcases}
\end{myequation}
along with the eight natural boundary conditions
 \begin{myequation}\label{A_nBCs}
    \begin{dcases}
        \frac{\partial \mathcal{M}}{\partial u_{1,2}}=0,~~~\quad\qquad\qquad \frac{\partial \mathcal{M}}{\partial u_{2,2}}=0, & \text{on} \quad   \partial\mathcal{ B}^+_m-\partial\mathcal{ B}_f \quad (X_2=H_m+H_f)\\
                \frac{\mu_m}{\mu_f}\left(\frac{\partial \mathcal{M}}{\partial u_{1,2}}\right)-\frac{\partial \mathcal{F}}{\partial u_{1,2}}=0, ~~\frac{\mu_m}{\mu_f}\left(\frac{\partial \mathcal{M}}{\partial u_{2,2}}\right)-\frac{\partial \mathcal{F}}{\partial u_{2,2}}=0, & \text{on} \quad   \partial\mathcal{ B}^+_m\cup\partial\mathcal{ B}_f \quad (X_2=H_f)\\
                                \frac{\mu_m}{\mu_f}\left(\frac{\partial \mathcal{M}}{\partial u_{1,2}}\right)-\frac{\partial \mathcal{F}}{\partial u_{1,2}}=0, ~~\frac{\mu_m}{\mu_f}\left(\frac{\partial \mathcal{M}}{\partial u_{2,2}}\right)-\frac{\partial \mathcal{F}}{\partial u_{2,2}}=0, & \text{on} \quad   \partial\mathcal{ B}^-_m\cup\partial\mathcal{ B}_f \quad (X_2=0)\\
        \frac{\partial \mathcal{M}}{\partial u_{1,2}}=0, ~~~\quad\qquad\qquad \frac{\partial \mathcal{M}}{\partial u_{2,2}}=0, & \text{on} \quad   \partial\mathcal{ B}^-_m-\partial\mathcal{ B}_f \quad (X_2=-H_m)
    \end{dcases}
\end{myequation}
By substituting \eqref{M_A} and \eqref{F_A} into the above equations, we arrive at the governing equations in \eqref{gov_eq} and the natural boundary conditions in \eqref{nBCs_free} and \eqref{nBCs_interface}.

\newpage
\noindent {\bf{The coefficients of matrix M:}}

  \begin{displaymath}\label{coeff}
  \begin{aligned}
  M_{1,1}&=(T_1^2+1)e^{T_1K\lambda ^{-1}(H_m\lambda_m ^{-1}+H_f\lambda_f ^{-1})},\\
  M_{1,2}&=(T_2^2+1)e^{T_2K\lambda ^{-1}(H_m\lambda_m ^{-1}+H_f\lambda_f ^{-1})},\\
  M_{1,3}&=(T_3^2+1)e^{T_3K\lambda ^{-1}(H_m\lambda_m ^{-1}+H_f\lambda_f ^{-1})},\\
  M_{1,4}&=(T_4^2+1)e^{T_4K\lambda ^{-1}(H_m\lambda_m ^{-1}+H_f\lambda_f ^{-1})},\\
  M_{1,5}&=\cdot \cdot \cdot =M_{1,12}=0,
  \end{aligned}   
  \qquad 
  \begin{aligned}
  M_{2,1}&=(-2\lambda_m^{-2}T_1+\lambda_m^{-2}T_1^3-\lambda_m^{2}T_1)e^{T_1K\lambda ^{-1}(H_m\lambda_m ^{-1}+H_f\lambda_f ^{-1})},\\
  M_{2,2}&=(-2\lambda_m^{-2}T_2+\lambda_m^{-2}T_2^3-\lambda_m^{2}T_2)e^{T_2K\lambda ^{-1}(H_m\lambda_m ^{-1}+H_f\lambda_f ^{-1})},\\
  M_{2,3}&=(-2\lambda_m^{-2}T_3+\lambda_m^{-2}T_3^3-\lambda_m^{2}T_3)e^{T_3K\lambda ^{-1}(H_m\lambda_m ^{-1}+H_f\lambda_f ^{-1})},\\
  M_{2,4}&=(-2\lambda_m^{-2}T_4+\lambda_m^{-2}T_4^3-\lambda_m^{2}T_4)e^{T_4K\lambda ^{-1}(H_m\lambda_m ^{-1}+H_f\lambda_f ^{-1})},\\
  M_{2,5}&=\cdot \cdot \cdot =M_{2,12}=0,
  \end{aligned}
  \end{displaymath}

 \bigskip

  \begin{displaymath}
  \begin{aligned}
  M_{3,1}&=-\lambda_m^{-2}(T_1^2+1)e^{T_1KH_f\lambda ^{-1}\lambda_f ^{-1}},\\
  M_{3,2}&=-\lambda_m^{-2}(T_2^2+1)e^{T_2KH_f\lambda ^{-1}\lambda_f ^{-1}},\\
  M_{3,3}&=-\lambda_m^{-2}(T_3^2+1)e^{T_3KH_f\lambda ^{-1}\lambda_f ^{-1}},\\
  M_{3,4}&=-\lambda_m^{-2}(T_4^2+1)e^{T_4KH_f\lambda ^{-1}\lambda_f ^{-1}},\\
  M_{3,5}&=\frac{\mu_f}{\mu_m}(C_{uyuy}t_1^2+C_{uyvx})e^{t_1KH_f\lambda ^{-1}\lambda_f ^{-1}},\\
  M_{3,6}&=\frac{\mu_f}{\mu_m}(C_{uyuy}t_2^2+C_{uyvx})e^{t_2KH_f\lambda ^{-1}\lambda_f ^{-1}},\\
  M_{3,7}&=\frac{\mu_f}{\mu_m}(C_{uyuy}t_3^2+C_{uyvx})e^{t_3KH_f\lambda ^{-1}\lambda_f ^{-1}},\\
  M_{3,8}&=\frac{\mu_f}{\mu_m}(C_{uyuy}t_4^2+C_{uyvx})e^{t_4KH_f\lambda ^{-1}\lambda_f ^{-1}},\\
  M_{3,9}&=\cdot \cdot \cdot =M_{3,12}=0,
 \end{aligned}   
  \qquad 
  \begin{aligned}
  M_{4,1} &= (-2\lambda_m^{-2}T_1+\lambda_m^{-2}T_1^3-\lambda_m^{2}T_1)e^{T_1KH_f\lambda ^{-1}\lambda_f ^{-1}},\\
  M_{4,2} &= (-2\lambda_m^{-2}T_2+\lambda_m^{-2}T_2^3-\lambda_m^{2}T_2)e^{T_2KH_f\lambda ^{-1}\lambda_f ^{-1}},\\
  M_{4,3} &= (-2\lambda_m^{-2}T_3+\lambda_m^{-2}T_3^3-\lambda_m^{2}T_3)e^{T_3KH_f\lambda ^{-1}\lambda_f ^{-1}},\\
  M_{4,4} &= (-2\lambda_m^{-2}T_4+\lambda_m^{-2}T_4^3-\lambda_m^{2}T_4)e^{T_4KH_f\lambda ^{-1}\lambda_f ^{-1}},\\
  M_{4,5} &= \frac{\mu_f}{\mu_m}[(C_{vyux}-C_{vyvy})t_1+C_{uyuy}t_1^3\qquad
  \\&~~~~+(C_{uxvy}+C_{uyvx}-C_{uxux})t_1]e^{t_1KH_f\lambda ^{-1}\lambda_f ^{-1}},
  \\
  M_{4,6}&=\frac{\mu_f}{\mu_m}[(C_{vyux}-C_{vyvy})t_2+C_{uyuy}t_2^3\qquad
  \\&~~~~+(C_{uxvy}+C_{uyvx}-C_{uxux})t_2]e^{t_2KH_f\lambda ^{-1}\lambda_f ^{-1}},
  \\
  M_{4,7}&=\frac{\mu_f}{\mu_m}[(C_{vyux}-C_{vyvy})t_3+C_{uyuy}t_3^3\qquad
  \\&~~~~+(C_{uxvy}+C_{uyvx}-C_{uxux})t_3]e^{t_3KH_f\lambda ^{-1}\lambda_f ^{-1}},
  \\
 M_{4,8}&=\frac{\mu_f}{\mu_m}[(C_{vyux}-C_{vyvy})t_4+C_{uyuy}t_4^3\qquad
 \\&~~~~+(C_{uxvy}+C_{uyvx}-C_{uxux})t_4]e^{t_4KH_f\lambda ^{-1}\lambda_f ^{-1}},
 \\
  M_{4,9}&=\cdot \cdot \cdot =M_{4,12}=0,
\end{aligned} 
\end{displaymath}
     
  \begin{displaymath}
  \begin{aligned}
  M_{5,1}&=-T_1e^{T_1KH_f\lambda ^{-1}\lambda_f ^{-1}},\\
  M_{5,2}&=-T_2e^{T_2KH_f\lambda ^{-1}\lambda_f ^{-1}},\\
  M_{5,3}&=-T_3e^{T_3KH_f\lambda ^{-1}\lambda_f ^{-1}},\\
  M_{5,4}&=-T_4e^{T_4KH_f\lambda ^{-1}\lambda_f ^{-1}},\\
  M_{5,5}&=t_1e^{t_1KH_f\lambda ^{-1}\lambda_f ^{-1}},\\
  M_{5,6}&=t_2e^{t_2KH_f\lambda ^{-1}\lambda_f ^{-1}},\\
  M_{5,7}&=t_3e^{t_3KH_f\lambda ^{-1}\lambda_f ^{-1}},\\
  M_{5,8}&=t_4e^{t_4KH_f\lambda ^{-1}\lambda_f ^{-1}},\\
  M_{5,9}&=\cdot \cdot \cdot =M_{5,12}=0,
   \end{aligned}   
  \qquad 
  \begin{aligned} 
  M_{6,1}&=-e^{T_1KH_f\lambda ^{-1}\lambda_f ^{-1}},\\
  M_{6,2}&=-e^{T_2KH_f\lambda ^{-1}\lambda_f ^{-1}},\\
  M_{6,3}&=-e^{T_3KH_f\lambda ^{-1}\lambda_f ^{-1}},\\
  M_{6,4}&=-e^{T_4KH_f\lambda ^{-1}\lambda_f ^{-1}},\\
  M_{6,5}&=-e^{t_1KH_f\lambda ^{-1}\lambda_f ^{-1}},\\
  M_{6,6}&=-e^{t_2KH_f\lambda ^{-1}\lambda_f ^{-1}},\\
  M_{6,7}&=-e^{t_3KH_f\lambda ^{-1}\lambda_f ^{-1}},\\
  M_{6,8}&=-e^{t_4KH_f\lambda ^{-1}\lambda_f ^{-1}},\\
  M_{6,9}&=\cdot \cdot \cdot =M_{6,12}=0,
  \end{aligned}   
  \qquad 
  \begin{aligned} 
  M_{7,1}&=\cdot \cdot \cdot =M_{7,4}=0,\\
  M_{7,5}&=\frac{\mu_f}{\mu_m}(C_{uyuy}t_1^2+C_{uyvx}),\\
  M_{7,6}&=\frac{\mu_f}{\mu_m}(C_{uyuy}t_2^2+C_{uyvx}),\\
  M_{7,7}&=\frac{\mu_f}{\mu_m}(C_{uyuy}t_3^2+C_{uyvx}),\\
  M_{7,8}&=\frac{\mu_f}{\mu_m}(C_{uyuy}t_4^2+C_{uyvx}),\\
  M_{7,9}&=-\lambda_m^{-2}(T_1^2+1),\\
  M_{7,10}&=-\lambda_m^{-2}(T_2^2+1),\\
  M_{7,11}&=-\lambda_m^{-2}(T_3^2+1),\\
  M_{7,12}&=-\lambda_m^{-2}(T_4^2+1),\\
  \end{aligned} 
\end{displaymath}
 
 \bigskip
   \begin{displaymath} 
    \begin{aligned} 
  M_{8,1}=\cdot \cdot \cdot =M_{8,4}=0,\\
   M_{8,5}=\frac{\mu_f}{\mu_m}[(C_{vyux}-C_{vyvy})t_1+C_{uyuy}t_1^3+(C_{uxvy}+C_{uyvx}-C_{uxux})t_1],\\
   M_{8,6}=\frac{\mu_f}{\mu_m}[(C_{vyux}-C_{vyvy})t_2+C_{uyuy}t_2^3+(C_{uxvy}+C_{uyvx}-C_{uxux})t_2],\\
   M_{8,7}=\frac{\mu_f}{\mu_m}[(C_{vyux}-C_{vyvy})t_3+C_{uyuy}t_3^3+(C_{uxvy}+C_{uyvx}-C_{uxux})t_3],\\
   M_{8,8}=\frac{\mu_f}{\mu_m}[(C_{vyux}-C_{vyvy})t_4+C_{uyuy}t_4^3+(C_{uxvy}+C_{uyvx}-C_{uxux})t_4],\\
   M_{8,9}=(-2\lambda_m^{-2}T_1+\lambda_m^{-2}T_1^3-\lambda_m^{2}T_1),\\
   M_{8,10}=(-2\lambda_m^{-2}T_2+\lambda_m^{-2}T_2^3-\lambda_m^{2}T_2),\\
   M_{8,11}=(-2\lambda_m^{-2}T_3+\lambda_m^{-2}T_3^3-\lambda_m^{2}T_3),\\
   M_{8,12}=(-2\lambda_m^{-2}T_4+\lambda_m^{-2}T_4^3-\lambda_m^{2}T_4),\\
     \end{aligned} 
\end{displaymath}
   
   \bigskip
    \begin{displaymath} 
    \begin{aligned}  
   M_{9,1}=\cdot \cdot \cdot =M_{9,4}=0,\\
   M_{9,5}=t_1,\\
   M_{9,6}=t_2,\\
   M_{9,7}=t_3,\\
   M_{9,8}=t_4,\\
   M_{9,9}=-T_1,\\
   M_{9,10}=-T_2,\\
   M_{9,11}=-T_3,\\
   M_{9,12}=-T_4,
    \end{aligned}   
  \qquad 
  \begin{aligned}  
   M_{10,1}=\cdot \cdot \cdot =M_{10,4}=0,\\
   M_{10,5}=\cdot \cdot \cdot =M_{10,7}=1,\\
   M_{10,8}=\cdot \cdot \cdot =M_{10,12}=-1,\\
~~\\
  M_{11,1}=\cdot \cdot \cdot =M_{11,8}=0,\\
  M_{11,9}=(T_1^2+1)e^{-T_1KH_m\lambda ^{-1}\lambda_m ^{-1}},\\
  M_{11,10}=(T_2^2+1)e^{-T_2KH_m\lambda ^{-1}\lambda_m ^{-1}},\\
  M_{11,11}=(T_3^2+1)e^{-T_3KH_m\lambda ^{-1}\lambda_m ^{-1}},\\
  M_{11,12}=(T_4^2+1)e^{-T_4KH_m\lambda ^{-1}\lambda_m ^{-1}},\\
   \end{aligned} 
\end{displaymath}

    \bigskip
\begin{displaymath} 
    \begin{aligned}  
  M_{12,1}=\cdot \cdot \cdot =M_{12,8}=0,\\
  M_{12,9}=(-2\lambda_m^{-2}T_1+\lambda_m^{-2}T_1^3-\lambda_m^{2}T_1)e^{-T_1KH_m\lambda ^{-1}\lambda_m ^{-1}},\\
  M_{12,10}=(-2\lambda_m^{-2}T_2+\lambda_m^{-2}T_2^3-\lambda_m^{2}T_2)e^{-T_2KH_m\lambda ^{-1}\lambda_m ^{-1}},\\
  M_{12,11}=(-2\lambda_m^{-2}T_3+\lambda_m^{-2}T_3^3-\lambda_m^{2}T_3)e^{-T_3KH_m\lambda ^{-1}\lambda_m ^{-1}},\\
  M_{12,12}=(-2\lambda_m^{-2}T_4+\lambda_m^{-2}T_4^3-\lambda_m^{2}T_4)e^{-T_4KH_m\lambda ^{-1}\lambda_m ^{-1}},
    \end{aligned} 
\end{displaymath}

\bigskip
  In the above formula $\lambda_f$ and $\lambda_m$ refer to the applied deformation in the film and matrix along the $x_1$ direction, namely, $\lambda_m=\lambda ,\lambda_f=\lambda_p \lambda$, and $t_1, t_2, t_3, t_4$ are obtained via equations \eqref{root}.
  Additionally, we have substituted 
     \begin{displaymath} 
  T_1=1,\quad T_2=\lambda_m^2,\quad T_3=-1, \quad T_4=\lambda_m^{-2} 
  \end{displaymath}
and 
  \begin{displaymath} \begin{aligned}
  C_{uxux}=\lambda_f^2-2\alpha(3\lambda_f^4-2\lambda_f^2+1),\\
    C_{uxvy}=C_{vyux}=\lambda_f^{-2}+2\alpha(3\lambda_f^{-4}-2\lambda_f^{-2}-1),\\
  C_{uyuy}=C_{uyvx}=\lambda_f^{-2}-2\alpha(\lambda_f^{-4}-2\lambda_f^{-2}+1),\\
  C_{vyvy}=\lambda_f^{-2}-2\alpha(3\lambda_f^{-4}-2\lambda_f^{-2}-1).
\end{aligned}\end{displaymath}

\newpage
\nolinenumbers
\bibliographystyle{elsarticle-harv}
\bibliography{ref}

\begin{thebibliography}{75}
\expandafter\ifx\csname natexlab\endcsname\relax\def\natexlab#1{#1}\fi
\providecommand{\url}[1]{\texttt{#1}}
\providecommand{\href}[2]{#2}
\providecommand{\path}[1]{#1}
\providecommand{\DOIprefix}{doi:}
\providecommand{\ArXivprefix}{arXiv:}
\providecommand{\URLprefix}{URL: }
\providecommand{\Pubmedprefix}{pmid:}
\providecommand{\doi}[1]{\href{http://dx.doi.org/#1}{\path{#1}}}
\providecommand{\Pubmed}[1]{\href{pmid:#1}{\path{#1}}}
\providecommand{\bibinfo}[2]{#2}
\ifx\xfnm\relax \def\xfnm[#1]{\unskip,\space#1}\fi
\bibitem[{Ashby et~al.(1989)Ashby, Blunt and Bannister}]{ashby1989flow}
\bibinfo{author}{Ashby, M.}, \bibinfo{author}{Blunt, F.},
  \bibinfo{author}{Bannister, M.}, \bibinfo{year}{1989}.
\newblock \bibinfo{title}{Flow characteristics of highly constrained metal
  wires}.
\newblock \bibinfo{journal}{Acta Metallurgica} \bibinfo{volume}{37},
  \bibinfo{pages}{1847--1857}.
\bibitem[{Audoly and Hutchinson(2016)}]{audoly2016analysis}
\bibinfo{author}{Audoly, B.}, \bibinfo{author}{Hutchinson, J.W.},
  \bibinfo{year}{2016}.
\newblock \bibinfo{title}{Analysis of necking based on a one-dimensional
  model}.
\newblock \bibinfo{journal}{Journal of the Mechanics and Physics of Solids}
  \bibinfo{volume}{97}, \bibinfo{pages}{68--91}.
\bibitem[{Audoly and Hutchinson(2019)}]{audoly2019one}
\bibinfo{author}{Audoly, B.}, \bibinfo{author}{Hutchinson, J.W.},
  \bibinfo{year}{2019}.
\newblock \bibinfo{title}{One-dimensional modeling of necking in rate-dependent
  materials}.
\newblock \bibinfo{journal}{Journal of the Mechanics and Physics of Solids}
  \bibinfo{volume}{123}, \bibinfo{pages}{149--171}.
\bibitem[{Ban et~al.(2019)Ban, Wang, Franklin, Liphardt, Janmey and
  Shenoy}]{ban2019strong}
\bibinfo{author}{Ban, E.}, \bibinfo{author}{Wang, H.},
  \bibinfo{author}{Franklin, J.M.}, \bibinfo{author}{Liphardt, J.T.},
  \bibinfo{author}{Janmey, P.A.}, \bibinfo{author}{Shenoy, V.B.},
  \bibinfo{year}{2019}.
\newblock \bibinfo{title}{Strong triaxial coupling and anomalous poisson effect
  in collagen networks}.
\newblock \bibinfo{journal}{Proceedings of the National Academy of Sciences}
  \bibinfo{volume}{116}, \bibinfo{pages}{6790--6799}.
\bibitem[{Banerjee et~al.(2015)Banerjee, Utuje and
  Marchetti}]{banerjee2015propagating}
\bibinfo{author}{Banerjee, S.}, \bibinfo{author}{Utuje, K.J.},
  \bibinfo{author}{Marchetti, M.C.}, \bibinfo{year}{2015}.
\newblock \bibinfo{title}{Propagating stress waves during epithelial
  expansion}.
\newblock \bibinfo{journal}{Physical review letters} \bibinfo{volume}{114},
  \bibinfo{pages}{228101}.
\bibitem[{Bertoldi and Boyce(2008)}]{bertoldi2008wave}
\bibinfo{author}{Bertoldi, K.}, \bibinfo{author}{Boyce, M.C.},
  \bibinfo{year}{2008}.
\newblock \bibinfo{title}{Wave propagation and instabilities in monolithic and
  periodically structured elastomeric materials undergoing large deformations}.
\newblock \bibinfo{journal}{Physical Review B} \bibinfo{volume}{78},
  \bibinfo{pages}{184107}.
\bibitem[{Biot(1963)}]{biot1963surface}
\bibinfo{author}{Biot, M.A.}, \bibinfo{year}{1963}.
\newblock \bibinfo{title}{Surface instability of rubber in compression}.
\newblock \bibinfo{journal}{Applied Scientific Research, Section A}
  \bibinfo{volume}{12}, \bibinfo{pages}{168--182}.
\bibitem[{Burla et~al.(2020)Burla, Dussi, Martinez-Torres, Tauber, van~der
  Gucht and Koenderink}]{burla2020connectivity}
\bibinfo{author}{Burla, F.}, \bibinfo{author}{Dussi, S.},
  \bibinfo{author}{Martinez-Torres, C.}, \bibinfo{author}{Tauber, J.},
  \bibinfo{author}{van~der Gucht, J.}, \bibinfo{author}{Koenderink, G.H.},
  \bibinfo{year}{2020}.
\newblock \bibinfo{title}{Connectivity and plasticity determine collagen
  network fracture}.
\newblock \bibinfo{journal}{Proceedings of the National Academy of Sciences}
  \bibinfo{volume}{117}, \bibinfo{pages}{8326--8334}.
\bibitem[{Cao and Hutchinson(2012)}]{cao2012wrinkles}
\bibinfo{author}{Cao, Y.}, \bibinfo{author}{Hutchinson, J.W.},
  \bibinfo{year}{2012}.
\newblock \bibinfo{title}{From wrinkles to creases in elastomers: the
  instability and imperfection-sensitivity of wrinkling}.
\newblock \bibinfo{journal}{Proceedings of the Royal Society A: Mathematical,
  Physical and Engineering Sciences} \bibinfo{volume}{468},
  \bibinfo{pages}{94--115}.
\bibitem[{Caporizzo and Prosser(2022)}]{caporizzo2022microtubule}
\bibinfo{author}{Caporizzo, M.A.}, \bibinfo{author}{Prosser, B.L.},
  \bibinfo{year}{2022}.
\newblock \bibinfo{title}{The microtubule cytoskeleton in cardiac mechanics and
  heart failure}.
\newblock \bibinfo{journal}{Nature Reviews Cardiology} \bibinfo{volume}{19},
  \bibinfo{pages}{364--378}.
\bibitem[{Cerik et~al.(2020)Cerik, Park and Choung}]{cerik2020use}
\bibinfo{author}{Cerik, B.C.}, \bibinfo{author}{Park, S.J.},
  \bibinfo{author}{Choung, J.}, \bibinfo{year}{2020}.
\newblock \bibinfo{title}{Use of localized necking and fracture as a failure
  criterion in ship collision analysis}.
\newblock \bibinfo{journal}{Marine Structures} \bibinfo{volume}{73},
  \bibinfo{pages}{102787}.
\bibitem[{Chaudhuri et~al.(2007)Chaudhuri, Parekh and
  Fletcher}]{chaudhuri2007reversible}
\bibinfo{author}{Chaudhuri, O.}, \bibinfo{author}{Parekh, S.H.},
  \bibinfo{author}{Fletcher, D.A.}, \bibinfo{year}{2007}.
\newblock \bibinfo{title}{Reversible stress softening of actin networks}.
\newblock \bibinfo{journal}{Nature} \bibinfo{volume}{445},
  \bibinfo{pages}{295--298}.
\bibitem[{Chen and Hutchinson(2004a)}]{chen2004family}
\bibinfo{author}{Chen, X.}, \bibinfo{author}{Hutchinson, J.W.},
  \bibinfo{year}{2004}a.
\newblock \bibinfo{title}{A family of herringbone patterns in thin films}.
\newblock \bibinfo{journal}{Scripta materialia} \bibinfo{volume}{50},
  \bibinfo{pages}{797--801}.
\bibitem[{Chen and Hutchinson(2004b)}]{chen2004herringbone}
\bibinfo{author}{Chen, X.}, \bibinfo{author}{Hutchinson, J.W.},
  \bibinfo{year}{2004}b.
\newblock \bibinfo{title}{Herringbone buckling patterns of compressed thin
  films on compliant substrates}.
\newblock \bibinfo{journal}{J. Appl. Mech.} \bibinfo{volume}{71},
  \bibinfo{pages}{597--603}.
\bibitem[{Chen et~al.(2017)Chen, Liao, Liu and Chen}]{chen2017helical}
\bibinfo{author}{Chen, Y.}, \bibinfo{author}{Liao, X.}, \bibinfo{author}{Liu,
  Y.}, \bibinfo{author}{Chen, X.}, \bibinfo{year}{2017}.
\newblock \bibinfo{title}{Helical buckling of wires embedded in a soft matrix
  under axial compression}.
\newblock \bibinfo{journal}{Extreme Mechanics Letters} \bibinfo{volume}{17},
  \bibinfo{pages}{71--76}.
\bibitem[{Cho et~al.(2016)Cho, Weaver, P{\"o}selt, in't Veld, Boyce and
  Rutledge}]{cho2016engineering}
\bibinfo{author}{Cho, H.}, \bibinfo{author}{Weaver, J.C.},
  \bibinfo{author}{P{\"o}selt, E.}, \bibinfo{author}{in't Veld, P.J.},
  \bibinfo{author}{Boyce, M.C.}, \bibinfo{author}{Rutledge, G.C.},
  \bibinfo{year}{2016}.
\newblock \bibinfo{title}{Engineering the mechanics of heterogeneous soft
  crystals}.
\newblock \bibinfo{journal}{Advanced Functional Materials}
  \bibinfo{volume}{26}, \bibinfo{pages}{6938--6949}.
\bibitem[{Chockalingam et~al.(2021)Chockalingam, Roth, Henzel and
  Cohen}]{chockalingam2021probing}
\bibinfo{author}{Chockalingam, S.}, \bibinfo{author}{Roth, C.},
  \bibinfo{author}{Henzel, T.}, \bibinfo{author}{Cohen, T.},
  \bibinfo{year}{2021}.
\newblock \bibinfo{title}{Probing local nonlinear viscoelastic properties in
  soft materials}.
\newblock \bibinfo{journal}{Journal of the Mechanics and Physics of Solids}
  \bibinfo{volume}{146}, \bibinfo{pages}{104172}.
\bibitem[{Colin and Holland(2019)}]{colin2019layer}
\bibinfo{author}{Colin, J.}, \bibinfo{author}{Holland, M.A.},
  \bibinfo{year}{2019}.
\newblock \bibinfo{title}{Layer wrinkling in an inhomogeneous matrix}.
\newblock \bibinfo{journal}{International Journal of Solids and Structures}
  \bibinfo{volume}{156}, \bibinfo{pages}{119--125}.
\bibitem[{Considère(1885)}]{Considerecrete1885}
\bibinfo{author}{Considère, A.}, \bibinfo{year}{1885}.
\newblock \bibinfo{title}{Annales des ponts et chaussées}.
\newblock \bibinfo{journal}{q} \bibinfo{volume}{9}, \bibinfo{pages}{574--775}.
\bibitem[{Cooper et~al.(2019)Cooper, Joshipura, Parekh, Norkett, Mailen,
  Miller, Genzer and Dickey}]{cooper2019toughening}
\bibinfo{author}{Cooper, C.B.}, \bibinfo{author}{Joshipura, I.D.},
  \bibinfo{author}{Parekh, D.P.}, \bibinfo{author}{Norkett, J.},
  \bibinfo{author}{Mailen, R.}, \bibinfo{author}{Miller, V.M.},
  \bibinfo{author}{Genzer, J.}, \bibinfo{author}{Dickey, M.D.},
  \bibinfo{year}{2019}.
\newblock \bibinfo{title}{Toughening stretchable fibers via serial fracturing
  of a metallic core}.
\newblock \bibinfo{journal}{Science advances} \bibinfo{volume}{5},
  \bibinfo{pages}{eaat4600}.
\bibitem[{Cui et~al.(2020)Cui, King, Huang, Chen, Sun, Guo, Saruwatari, Hui,
  Kurokawa and Gong}]{cui2020fiber}
\bibinfo{author}{Cui, W.}, \bibinfo{author}{King, D.R.},
  \bibinfo{author}{Huang, Y.}, \bibinfo{author}{Chen, L.},
  \bibinfo{author}{Sun, T.L.}, \bibinfo{author}{Guo, Y.},
  \bibinfo{author}{Saruwatari, Y.}, \bibinfo{author}{Hui, C.Y.},
  \bibinfo{author}{Kurokawa, T.}, \bibinfo{author}{Gong, J.P.},
  \bibinfo{year}{2020}.
\newblock \bibinfo{title}{Fiber-reinforced viscoelastomers show extraordinary
  crack resistance that exceeds metals}.
\newblock \bibinfo{journal}{Advanced Materials} \bibinfo{volume}{32},
  \bibinfo{pages}{1907180}.
\bibitem[{Daly et~al.(2021)Daly, Prendergast, Hughes and
  Burdick}]{daly2021bioprinting}
\bibinfo{author}{Daly, A.C.}, \bibinfo{author}{Prendergast, M.E.},
  \bibinfo{author}{Hughes, A.J.}, \bibinfo{author}{Burdick, J.A.},
  \bibinfo{year}{2021}.
\newblock \bibinfo{title}{Bioprinting for the biologist}.
\newblock \bibinfo{journal}{Cell} \bibinfo{volume}{184},
  \bibinfo{pages}{18--32}.
\bibitem[{Dortdivanlioglu and Javili(2022)}]{dortdivanlioglu2022plateau}
\bibinfo{author}{Dortdivanlioglu, B.}, \bibinfo{author}{Javili, A.},
  \bibinfo{year}{2022}.
\newblock \bibinfo{title}{Plateau rayleigh instability of soft elastic solids.
  effect of compressibility on pre and post bifurcation behavior}.
\newblock \bibinfo{journal}{Extreme Mechanics Letters} ,
  \bibinfo{pages}{101797}.
\bibitem[{Douville et~al.(2011)Douville, Li, Takayama and
  Thouless}]{douville2011fracture}
\bibinfo{author}{Douville, N.J.}, \bibinfo{author}{Li, Z.},
  \bibinfo{author}{Takayama, S.}, \bibinfo{author}{Thouless, M.},
  \bibinfo{year}{2011}.
\newblock \bibinfo{title}{Fracture of metal coated elastomers}.
\newblock \bibinfo{journal}{Soft Matter} \bibinfo{volume}{7},
  \bibinfo{pages}{6493--6500}.
\bibitem[{Du et~al.(2019)Du, L{\"u}, Liu, Han, Li, Chen, Qu and
  Destrade}]{du2019prescribing}
\bibinfo{author}{Du, Y.}, \bibinfo{author}{L{\"u}, C.}, \bibinfo{author}{Liu,
  C.}, \bibinfo{author}{Han, Z.}, \bibinfo{author}{Li, J.},
  \bibinfo{author}{Chen, W.}, \bibinfo{author}{Qu, S.},
  \bibinfo{author}{Destrade, M.}, \bibinfo{year}{2019}.
\newblock \bibinfo{title}{Prescribing patterns in growing tubular soft matter
  by initial residual stress}.
\newblock \bibinfo{journal}{Soft matter} \bibinfo{volume}{15},
  \bibinfo{pages}{8468--8474}.
\bibitem[{Fu et~al.(2021)Fu, Jin and Goriely}]{fu2021necking}
\bibinfo{author}{Fu, Y.}, \bibinfo{author}{Jin, L.}, \bibinfo{author}{Goriely,
  A.}, \bibinfo{year}{2021}.
\newblock \bibinfo{title}{Necking, beading, and bulging in soft elastic
  cylinders}.
\newblock \bibinfo{journal}{Journal of the Mechanics and Physics of Solids}
  \bibinfo{volume}{147}, \bibinfo{pages}{104250}.
\bibitem[{Gao and Li(2017)}]{gao2017tuning}
\bibinfo{author}{Gao, C.}, \bibinfo{author}{Li, Y.}, \bibinfo{year}{2017}.
\newblock \bibinfo{title}{Tuning the wrinkling patterns of an
  interfacial/coating layer via a regulation interphase}.
\newblock \bibinfo{journal}{International Journal of Solids and Structures}
  \bibinfo{volume}{104}, \bibinfo{pages}{92--102}.
\bibitem[{Goehring(2013)}]{goehring2013evolving}
\bibinfo{author}{Goehring, L.}, \bibinfo{year}{2013}.
\newblock \bibinfo{title}{Evolving fracture patterns: columnar joints, mud
  cracks and polygonal terrain}.
\newblock \bibinfo{journal}{Philosophical Transactions of the Royal Society A:
  Mathematical, Physical and Engineering Sciences} \bibinfo{volume}{371},
  \bibinfo{pages}{20120353}.
\bibitem[{Henzel et~al.(2022)Henzel, Nijjer, Chockalingam, Wahdat, Crosby, Yan
  and Cohen}]{HenzelPNASn}
\bibinfo{author}{Henzel, T.}, \bibinfo{author}{Nijjer, J.},
  \bibinfo{author}{Chockalingam, S.}, \bibinfo{author}{Wahdat, H.},
  \bibinfo{author}{Crosby, A.J.}, \bibinfo{author}{Yan, J.},
  \bibinfo{author}{Cohen, T.}, \bibinfo{year}{2022}.
\newblock \bibinfo{title}{{Interfacial cavitation}}.
\newblock \bibinfo{journal}{PNAS Nexus}
  \DOIprefix\doi{10.1093/pnasnexus/pgac217}.
\bibitem[{Herrmann et~al.(1967)Herrmann, Mason and Chan}]{herrmann1967response}
\bibinfo{author}{Herrmann, L.R.}, \bibinfo{author}{Mason, W.E.},
  \bibinfo{author}{Chan, S.}, \bibinfo{year}{1967}.
\newblock \bibinfo{title}{Response of reinforcing wires to compressive states
  of stress}.
\newblock \bibinfo{journal}{Journal of Composite Materials}
  \bibinfo{volume}{1}, \bibinfo{pages}{212--226}.
\bibitem[{Hill(1952)}]{hill1952discontinuous}
\bibinfo{author}{Hill, R.}, \bibinfo{year}{1952}.
\newblock \bibinfo{title}{On discontinuous plastic states, with special
  reference to localized necking in thin sheets}.
\newblock \bibinfo{journal}{Journal of the Mechanics and Physics of Solids}
  \bibinfo{volume}{1}, \bibinfo{pages}{19--30}.
\bibitem[{Holland et~al.(2017)Holland, Li, Feng and
  Kuhl}]{holland2017instabilities}
\bibinfo{author}{Holland, M.}, \bibinfo{author}{Li, B.}, \bibinfo{author}{Feng,
  X.}, \bibinfo{author}{Kuhl, E.}, \bibinfo{year}{2017}.
\newblock \bibinfo{title}{Instabilities of soft films on compliant substrates}.
\newblock \bibinfo{journal}{Journal of the Mechanics and Physics of Solids}
  \bibinfo{volume}{98}, \bibinfo{pages}{350--365}.
\bibitem[{Huang et~al.(2019)Huang, King, Cui, Sun, Guo, Kurokawa, Brown, Hui
  and Gong}]{huang2019superior}
\bibinfo{author}{Huang, Y.}, \bibinfo{author}{King, D.R.},
  \bibinfo{author}{Cui, W.}, \bibinfo{author}{Sun, T.L.}, \bibinfo{author}{Guo,
  H.}, \bibinfo{author}{Kurokawa, T.}, \bibinfo{author}{Brown, H.R.},
  \bibinfo{author}{Hui, C.Y.}, \bibinfo{author}{Gong, J.P.},
  \bibinfo{year}{2019}.
\newblock \bibinfo{title}{Superior fracture resistance of fiber reinforced
  polyampholyte hydrogels achieved by extraordinarily large energy-dissipative
  process zones}.
\newblock \bibinfo{journal}{Journal of Materials Chemistry A}
  \bibinfo{volume}{7}, \bibinfo{pages}{13431--13440}.
\bibitem[{Hutchinson and Neale(1978)}]{hutchinson1978sheet}
\bibinfo{author}{Hutchinson, J.}, \bibinfo{author}{Neale, K.},
  \bibinfo{year}{1978}.
\newblock \bibinfo{title}{Sheet necking-ii. time-independent behavior}, in:
  \bibinfo{booktitle}{Mechanics of sheet metal forming}.
  \bibinfo{publisher}{Springer}, pp. \bibinfo{pages}{127--153}.
\bibitem[{Jia and Li(2019)}]{jia2019bifurcation}
\bibinfo{author}{Jia, Z.}, \bibinfo{author}{Li, T.}, \bibinfo{year}{2019}.
\newblock \bibinfo{title}{Bifurcation instability of substrate-supported metal
  films under biaxial in-plane tension}.
\newblock \bibinfo{journal}{Journal of the Mechanics and Physics of Solids}
  \bibinfo{volume}{126}, \bibinfo{pages}{52--75}.
\bibitem[{Jia and Li(2021)}]{jia2021effect}
\bibinfo{author}{Jia, Z.}, \bibinfo{author}{Li, T.}, \bibinfo{year}{2021}.
\newblock \bibinfo{title}{Effect of interfacial stiffness on the stretchability
  of metal/elastomer bilayers under in-plane biaxial tension}.
\newblock \bibinfo{journal}{Theoretical and Applied Mechanics Letters}
  \bibinfo{volume}{11}, \bibinfo{pages}{100247}.
\bibitem[{Johnson et~al.(2013)Johnson, Stewart and Smith}]{JOHNSON201335}
\bibinfo{author}{Johnson, V.E.}, \bibinfo{author}{Stewart, W.},
  \bibinfo{author}{Smith, D.H.}, \bibinfo{year}{2013}.
\newblock \bibinfo{title}{Axonal pathology in traumatic brain injury}.
\newblock \bibinfo{journal}{Experimental neurology} \bibinfo{volume}{246},
  \bibinfo{pages}{35--43}.
\bibitem[{Kabir et~al.(2020)Kabir, Mathur and Seyam}]{kabir2020critical}
\bibinfo{author}{Kabir, S.F.}, \bibinfo{author}{Mathur, K.},
  \bibinfo{author}{Seyam, A.F.M.}, \bibinfo{year}{2020}.
\newblock \bibinfo{title}{A critical review on 3d printed continuous
  fiber-reinforced composites: History, mechanism, materials and properties}.
\newblock \bibinfo{journal}{Composite Structures} \bibinfo{volume}{232},
  \bibinfo{pages}{111476}.
\bibitem[{Kothari et~al.(2020)Kothari, Lemon, Roth and
  Cohen}]{kothari2020controlled}
\bibinfo{author}{Kothari, M.}, \bibinfo{author}{Lemon, Z.},
  \bibinfo{author}{Roth, C.}, \bibinfo{author}{Cohen, T.},
  \bibinfo{year}{2020}.
\newblock \bibinfo{title}{Controlled propagation and jamming of a delamination
  front}.
\newblock \bibinfo{journal}{Soft Matter} \bibinfo{volume}{16},
  \bibinfo{pages}{9838--9843}.
\bibitem[{Kravchenko et~al.(2016)Kravchenko, Kravchenko and
  Pipes}]{kravchenko2016chemical}
\bibinfo{author}{Kravchenko, O.G.}, \bibinfo{author}{Kravchenko, S.G.},
  \bibinfo{author}{Pipes, R.B.}, \bibinfo{year}{2016}.
\newblock \bibinfo{title}{Chemical and thermal shrinkage in thermosetting
  prepreg}.
\newblock \bibinfo{journal}{Composites Part A: Applied Science and
  Manufacturing} \bibinfo{volume}{80}, \bibinfo{pages}{72--81}.
\bibitem[{Leonov(2002)}]{leonov2002theory}
\bibinfo{author}{Leonov, A.}, \bibinfo{year}{2002}.
\newblock \bibinfo{title}{A theory of necking in semi-crystalline polymers}.
\newblock \bibinfo{journal}{International journal of solids and structures}
  \bibinfo{volume}{39}, \bibinfo{pages}{5913--5926}.
\bibitem[{Li et~al.(2019)Li, Pallicity, Slesarenko, Goshkoderia and
  Rudykh}]{li2019domain}
\bibinfo{author}{Li, J.}, \bibinfo{author}{Pallicity, T.D.},
  \bibinfo{author}{Slesarenko, V.}, \bibinfo{author}{Goshkoderia, A.},
  \bibinfo{author}{Rudykh, S.}, \bibinfo{year}{2019}.
\newblock \bibinfo{title}{Domain formations and pattern transitions via
  instabilities in soft heterogeneous materials}.
\newblock \bibinfo{journal}{Advanced Materials} \bibinfo{volume}{31},
  \bibinfo{pages}{1807309}.
\bibitem[{Li et~al.(2018)Li, Slesarenko, Galich and
  Rudykh}]{li2018instabilities}
\bibinfo{author}{Li, J.}, \bibinfo{author}{Slesarenko, V.},
  \bibinfo{author}{Galich, P.I.}, \bibinfo{author}{Rudykh, S.},
  \bibinfo{year}{2018}.
\newblock \bibinfo{title}{Instabilities and pattern formations in 3d-printed
  deformable fiber composites}.
\newblock \bibinfo{journal}{Composites Part B: Engineering}
  \bibinfo{volume}{148}, \bibinfo{pages}{114--122}.
\bibitem[{Li et~al.(2022a)Li, Slesarenko and Rudykh}]{li2022emergence}
\bibinfo{author}{Li, J.}, \bibinfo{author}{Slesarenko, V.},
  \bibinfo{author}{Rudykh, S.}, \bibinfo{year}{2022}a.
\newblock \bibinfo{title}{Emergence of instability-driven domains in soft
  stratified materials}.
\newblock \bibinfo{journal}{npj Computational Materials} \bibinfo{volume}{8},
  \bibinfo{pages}{1--6}.
\bibitem[{Li et~al.(2022b)Li, Yan, Xu, Wang, Wu and Feng}]{li2022surface}
\bibinfo{author}{Li, M.}, \bibinfo{author}{Yan, Y.}, \bibinfo{author}{Xu, S.},
  \bibinfo{author}{Wang, G.}, \bibinfo{author}{Wu, J.}, \bibinfo{author}{Feng,
  X.Q.}, \bibinfo{year}{2022}b.
\newblock \bibinfo{title}{Surface effect on the necking of hyperelastic
  materials}.
\newblock \bibinfo{journal}{Current Applied Physics} \bibinfo{volume}{38},
  \bibinfo{pages}{91--98}.
\bibitem[{Li and Suo(2006)}]{li2006deformability}
\bibinfo{author}{Li, T.}, \bibinfo{author}{Suo, Z.}, \bibinfo{year}{2006}.
\newblock \bibinfo{title}{Deformability of thin metal films on elastomer
  substrates}.
\newblock \bibinfo{journal}{International Journal of Solids and Structures}
  \bibinfo{volume}{43}, \bibinfo{pages}{2351--2363}.
\bibitem[{Lin et~al.(2016)Lin, Cohen, Zhang, Yuk, Abeyaratne and
  Zhao}]{lin2016fringe}
\bibinfo{author}{Lin, S.}, \bibinfo{author}{Cohen, T.}, \bibinfo{author}{Zhang,
  T.}, \bibinfo{author}{Yuk, H.}, \bibinfo{author}{Abeyaratne, R.},
  \bibinfo{author}{Zhao, X.}, \bibinfo{year}{2016}.
\newblock \bibinfo{title}{Fringe instability in constrained soft elastic
  layers}.
\newblock \bibinfo{journal}{Soft matter} \bibinfo{volume}{12},
  \bibinfo{pages}{8899--8906}.
\bibitem[{Liu et~al.(2020a)Liu, Du, L{\"u} and Chen}]{liu2020growth}
\bibinfo{author}{Liu, C.}, \bibinfo{author}{Du, Y.}, \bibinfo{author}{L{\"u},
  C.}, \bibinfo{author}{Chen, W.}, \bibinfo{year}{2020}a.
\newblock \bibinfo{title}{Growth and patterns of residually stressed
  core--shell soft sphere}.
\newblock \bibinfo{journal}{International Journal of Non-Linear Mechanics}
  \bibinfo{volume}{127}, \bibinfo{pages}{103594}.
\bibitem[{Liu et~al.(2020b)Liu, Li, Jiang, Jia, Xu and Wang}]{liu2020effect}
\bibinfo{author}{Liu, F.}, \bibinfo{author}{Li, T.}, \bibinfo{author}{Jiang,
  X.}, \bibinfo{author}{Jia, Z.}, \bibinfo{author}{Xu, Z.},
  \bibinfo{author}{Wang, L.}, \bibinfo{year}{2020}b.
\newblock \bibinfo{title}{The effect of material mixing on interfacial
  stiffness and strength of multi-material additive manufacturing}.
\newblock \bibinfo{journal}{Additive Manufacturing} \bibinfo{volume}{36},
  \bibinfo{pages}{101502}.
\bibitem[{Lwin et~al.(2022)Lwin, Sindermann, Sutter, Jackson, Bonassar, Cohen
  and Das}]{lwin2022rigidity}
\bibinfo{author}{Lwin, P.}, \bibinfo{author}{Sindermann, A.},
  \bibinfo{author}{Sutter, L.}, \bibinfo{author}{Jackson, T.W.},
  \bibinfo{author}{Bonassar, L.}, \bibinfo{author}{Cohen, I.},
  \bibinfo{author}{Das, M.}, \bibinfo{year}{2022}.
\newblock \bibinfo{title}{Rigidity and fracture of biopolymer double networks}.
\newblock \bibinfo{journal}{Soft Matter} \bibinfo{volume}{18},
  \bibinfo{pages}{322--327}.
\bibitem[{Matis(2020)}]{matis2020mechanical}
\bibinfo{author}{Matis, M.}, \bibinfo{year}{2020}.
\newblock \bibinfo{title}{The mechanical role of microtubules in tissue
  remodeling}.
\newblock \bibinfo{journal}{BioEssays} \bibinfo{volume}{42},
  \bibinfo{pages}{1900244}.
\bibitem[{Mora et~al.(2010)Mora, Phou, Fromental, Pismen and
  Pomeau}]{mora2010capillarity}
\bibinfo{author}{Mora, S.}, \bibinfo{author}{Phou, T.},
  \bibinfo{author}{Fromental, J.M.}, \bibinfo{author}{Pismen, L.M.},
  \bibinfo{author}{Pomeau, Y.}, \bibinfo{year}{2010}.
\newblock \bibinfo{title}{Capillarity driven instability of a soft solid}.
\newblock \bibinfo{journal}{Physical review letters} \bibinfo{volume}{105},
  \bibinfo{pages}{214301}.
\bibitem[{Morley et~al.(2019)Morley, Ellison, Bhattacharjee, O’Bryan, Zhang,
  Smith, Kabb, Sebastian, Moore, Schulze et~al.}]{morley2019quantitative}
\bibinfo{author}{Morley, C.D.}, \bibinfo{author}{Ellison, S.},
  \bibinfo{author}{Bhattacharjee, T.}, \bibinfo{author}{O’Bryan, C.S.},
  \bibinfo{author}{Zhang, Y.}, \bibinfo{author}{Smith, K.F.},
  \bibinfo{author}{Kabb, C.P.}, \bibinfo{author}{Sebastian, M.},
  \bibinfo{author}{Moore, G.L.}, \bibinfo{author}{Schulze, K.D.}, et~al.,
  \bibinfo{year}{2019}.
\newblock \bibinfo{title}{Quantitative characterization of 3d bioprinted
  structural elements under cell generated forces}.
\newblock \bibinfo{journal}{Nature communications} \bibinfo{volume}{10},
  \bibinfo{pages}{1--9}.
\bibitem[{Morovati et~al.(2020)Morovati, Saadat and
  Dargazany}]{morovati2020necking}
\bibinfo{author}{Morovati, V.}, \bibinfo{author}{Saadat, M.A.},
  \bibinfo{author}{Dargazany, R.}, \bibinfo{year}{2020}.
\newblock \bibinfo{title}{Necking of double-network gels: Constitutive modeling
  with microstructural insight}.
\newblock \bibinfo{journal}{Physical Review E} \bibinfo{volume}{102},
  \bibinfo{pages}{062501}.
\bibitem[{Nardinocchi and Puntel(2017)}]{nardinocchi2017swelling}
\bibinfo{author}{Nardinocchi, P.}, \bibinfo{author}{Puntel, E.},
  \bibinfo{year}{2017}.
\newblock \bibinfo{title}{Swelling-induced wrinkling in layered gel beams}.
\newblock \bibinfo{journal}{Proceedings of the Royal Society A: Mathematical,
  Physical and Engineering Sciences} \bibinfo{volume}{473},
  \bibinfo{pages}{20170454}.
\bibitem[{Raayai-Ardakani et~al.(2019)Raayai-Ardakani, Chen, Earl and
  Cohen}]{raayai2019volume}
\bibinfo{author}{Raayai-Ardakani, S.}, \bibinfo{author}{Chen, Z.},
  \bibinfo{author}{Earl, D.R.}, \bibinfo{author}{Cohen, T.},
  \bibinfo{year}{2019}.
\newblock \bibinfo{title}{Volume-controlled cavity expansion for probing of
  local elastic properties in soft materials}.
\newblock \bibinfo{journal}{Soft matter} \bibinfo{volume}{15},
  \bibinfo{pages}{381--392}.
\bibitem[{Riccobelli(2021)}]{riccobelli2021active}
\bibinfo{author}{Riccobelli, D.}, \bibinfo{year}{2021}.
\newblock \bibinfo{title}{Active elasticity drives the formation of periodic
  beading in damaged axons}.
\newblock \bibinfo{journal}{Physical Review E} \bibinfo{volume}{104},
  \bibinfo{pages}{024417}.
\bibitem[{Ringoot et~al.(2021)Ringoot, Roch, Molinari, Massart and
  Cohen}]{ringoot2021stick}
\bibinfo{author}{Ringoot, E.}, \bibinfo{author}{Roch, T.},
  \bibinfo{author}{Molinari, J.F.}, \bibinfo{author}{Massart, T.J.},
  \bibinfo{author}{Cohen, T.}, \bibinfo{year}{2021}.
\newblock \bibinfo{title}{Stick--slip phenomena and schallamach waves captured
  using reversible cohesive elements}.
\newblock \bibinfo{journal}{Journal of the Mechanics and Physics of Solids}
  \bibinfo{volume}{155}, \bibinfo{pages}{104528}.
\bibitem[{Rodr{\'\i}guez-Mart{\'\i}nez
  et~al.(2017)Rodr{\'\i}guez-Mart{\'\i}nez, Molinari, Zaera, Vadillo and
  Fern{\'a}ndez-S{\'a}ez}]{rodriguez2017critical}
\bibinfo{author}{Rodr{\'\i}guez-Mart{\'\i}nez, J.A.},
  \bibinfo{author}{Molinari, A.}, \bibinfo{author}{Zaera, R.},
  \bibinfo{author}{Vadillo, G.}, \bibinfo{author}{Fern{\'a}ndez-S{\'a}ez, J.},
  \bibinfo{year}{2017}.
\newblock \bibinfo{title}{The critical neck spacing in ductile plates subjected
  to dynamic biaxial loading: on the interplay between loading path and inertia
  effects}.
\newblock \bibinfo{journal}{International journal of solids and structures}
  \bibinfo{volume}{108}, \bibinfo{pages}{74--84}.
\bibitem[{de~Sa et~al.(2008)de~Sa, Benboudjema, Thiery and
  Sicard}]{de2008analysis}
\bibinfo{author}{de~Sa, C.}, \bibinfo{author}{Benboudjema, F.},
  \bibinfo{author}{Thiery, M.}, \bibinfo{author}{Sicard, J.},
  \bibinfo{year}{2008}.
\newblock \bibinfo{title}{Analysis of microcracking induced by differential
  drying shrinkage}.
\newblock \bibinfo{journal}{Cement and Concrete Composites}
  \bibinfo{volume}{30}, \bibinfo{pages}{947--956}.
\bibitem[{Saba and Jawaid(2018)}]{saba2018review}
\bibinfo{author}{Saba, N.}, \bibinfo{author}{Jawaid, M.}, \bibinfo{year}{2018}.
\newblock \bibinfo{title}{A review on thermomechanical properties of polymers
  and fibers reinforced polymer composites}.
\newblock \bibinfo{journal}{Journal of industrial and engineering chemistry}
  \bibinfo{volume}{67}, \bibinfo{pages}{1--11}.
\bibitem[{Shabahang et~al.(2016)Shabahang, Tao, Kaufman, Qiao, Wei, Bouchenot,
  Gordon, Fink, Bai, Hoy et~al.}]{shabahang2016controlled}
\bibinfo{author}{Shabahang, S.}, \bibinfo{author}{Tao, G.},
  \bibinfo{author}{Kaufman, J.J.}, \bibinfo{author}{Qiao, Y.},
  \bibinfo{author}{Wei, L.}, \bibinfo{author}{Bouchenot, T.},
  \bibinfo{author}{Gordon, A.P.}, \bibinfo{author}{Fink, Y.},
  \bibinfo{author}{Bai, Y.}, \bibinfo{author}{Hoy, R.S.}, et~al.,
  \bibinfo{year}{2016}.
\newblock \bibinfo{title}{Controlled fragmentation of multimaterial fibres and
  films via polymer cold-drawing}.
\newblock \bibinfo{journal}{Nature} \bibinfo{volume}{534},
  \bibinfo{pages}{529--533}.
\bibitem[{Slesarenko and Rudykh(2017)}]{slesarenko2017microscopic}
\bibinfo{author}{Slesarenko, V.}, \bibinfo{author}{Rudykh, S.},
  \bibinfo{year}{2017}.
\newblock \bibinfo{title}{Microscopic and macroscopic instabilities in
  hyperelastic fiber composites}.
\newblock \bibinfo{journal}{Journal of the Mechanics and Physics of Solids}
  \bibinfo{volume}{99}, \bibinfo{pages}{471--482}.
\bibitem[{Steif(1986)}]{steif1986periodic}
\bibinfo{author}{Steif, P.S.}, \bibinfo{year}{1986}.
\newblock \bibinfo{title}{Periodic necking instabilities in layered plastic
  composites}.
\newblock \bibinfo{journal}{International journal of solids and structures}
  \bibinfo{volume}{22}, \bibinfo{pages}{1571--1578}.
\bibitem[{St{\"o}ren and Rice(1975)}]{storen1975localized}
\bibinfo{author}{St{\"o}ren, S.}, \bibinfo{author}{Rice, J.},
  \bibinfo{year}{1975}.
\newblock \bibinfo{title}{Localized necking in thin sheets}.
\newblock \bibinfo{journal}{Journal of the Mechanics and Physics of Solids}
  \bibinfo{volume}{23}, \bibinfo{pages}{421--441}.
\bibitem[{Su et~al.(2014)Su, Liu, Terwagne, Reis and Bertoldi}]{su2014buckling}
\bibinfo{author}{Su, T.}, \bibinfo{author}{Liu, J.}, \bibinfo{author}{Terwagne,
  D.}, \bibinfo{author}{Reis, P.M.}, \bibinfo{author}{Bertoldi, K.},
  \bibinfo{year}{2014}.
\newblock \bibinfo{title}{Buckling of an elastic rod embedded on an elastomeric
  matrix: planar vs. non-planar configurations}.
\newblock \bibinfo{journal}{Soft Matter} \bibinfo{volume}{10},
  \bibinfo{pages}{6294--6302}.
\bibitem[{Thouless et~al.(2011)Thouless, Li, Douville and
  Takayama}]{thouless2011periodic}
\bibinfo{author}{Thouless, M.}, \bibinfo{author}{Li, Z.},
  \bibinfo{author}{Douville, N.}, \bibinfo{author}{Takayama, S.},
  \bibinfo{year}{2011}.
\newblock \bibinfo{title}{Periodic cracking of films supported on compliant
  substrates}.
\newblock \bibinfo{journal}{Journal of the Mechanics and Physics of Solids}
  \bibinfo{volume}{59}, \bibinfo{pages}{1927--1937}.
\bibitem[{Wang et~al.(2013)Wang, Svoronos, Boudou, Sakar, Schell, Morgan, Chen
  and Shenoy}]{wang2013necking}
\bibinfo{author}{Wang, H.}, \bibinfo{author}{Svoronos, A.A.},
  \bibinfo{author}{Boudou, T.}, \bibinfo{author}{Sakar, M.S.},
  \bibinfo{author}{Schell, J.Y.}, \bibinfo{author}{Morgan, J.R.},
  \bibinfo{author}{Chen, C.S.}, \bibinfo{author}{Shenoy, V.B.},
  \bibinfo{year}{2013}.
\newblock \bibinfo{title}{Necking and failure of constrained 3d microtissues
  induced by cellular tension}.
\newblock \bibinfo{journal}{Proceedings of the National Academy of Sciences}
  \bibinfo{volume}{110}, \bibinfo{pages}{20923--20928}.
\bibitem[{Xue and Hutchinson(2007)}]{xue2007neck}
\bibinfo{author}{Xue, Z.}, \bibinfo{author}{Hutchinson, J.W.},
  \bibinfo{year}{2007}.
\newblock \bibinfo{title}{Neck retardation and enhanced energy absorption in
  metal--elastomer bilayers}.
\newblock \bibinfo{journal}{Mechanics of Materials} \bibinfo{volume}{39},
  \bibinfo{pages}{473--487}.
\bibitem[{Xue and Hutchinson(2008)}]{xue2008neck}
\bibinfo{author}{Xue, Z.}, \bibinfo{author}{Hutchinson, J.W.},
  \bibinfo{year}{2008}.
\newblock \bibinfo{title}{Neck development in metal/elastomer bilayers under
  dynamic stretchings}.
\newblock \bibinfo{journal}{International Journal of Solids and Structures}
  \bibinfo{volume}{45}, \bibinfo{pages}{3769--3778}.
\bibitem[{Yan et~al.(2022)Yan, Li, Zhao and Feng}]{yan2022an}
\bibinfo{author}{Yan, L.}, \bibinfo{author}{Li, M.}, \bibinfo{author}{Zhao,
  Z.}, \bibinfo{author}{Feng, X.Q.}, \bibinfo{year}{2022}.
\newblock \bibinfo{title}{An energy method for the bifurcation analysis of
  necking}.
\newblock \bibinfo{journal}{Extreme Mechanics Letters} \bibinfo{volume}{55},
  \bibinfo{pages}{101793}.
\bibitem[{Yan et~al.(2020)Yan, Dong, Xiang, Jiang, Leber, Loke, Xu, Hou, Zhou,
  Chen et~al.}]{yan2020thermally}
\bibinfo{author}{Yan, W.}, \bibinfo{author}{Dong, C.}, \bibinfo{author}{Xiang,
  Y.}, \bibinfo{author}{Jiang, S.}, \bibinfo{author}{Leber, A.},
  \bibinfo{author}{Loke, G.}, \bibinfo{author}{Xu, W.}, \bibinfo{author}{Hou,
  C.}, \bibinfo{author}{Zhou, S.}, \bibinfo{author}{Chen, M.}, et~al.,
  \bibinfo{year}{2020}.
\newblock \bibinfo{title}{Thermally drawn advanced functional fibers: New
  frontier of flexible electronics}.
\newblock \bibinfo{journal}{Materials Today} \bibinfo{volume}{35},
  \bibinfo{pages}{168--194}.
\bibitem[{Zhao et~al.(2016)Zhao, Li, Cao and Feng}]{zhao2016buckling}
\bibinfo{author}{Zhao, Y.}, \bibinfo{author}{Li, J.}, \bibinfo{author}{Cao,
  Y.P.}, \bibinfo{author}{Feng, X.Q.}, \bibinfo{year}{2016}.
\newblock \bibinfo{title}{Buckling of an elastic fiber with finite length in a
  soft matrix}.
\newblock \bibinfo{journal}{Soft Matter} \bibinfo{volume}{12},
  \bibinfo{pages}{2086--2094}.
\bibitem[{Zhu et~al.(2005)Zhu, Mills, Peters, Bahng, Liu, Shim, Naruse, Csete,
  Thouless and Takayama}]{zhu2005fabrication}
\bibinfo{author}{Zhu, X.}, \bibinfo{author}{Mills, K.L.},
  \bibinfo{author}{Peters, P.R.}, \bibinfo{author}{Bahng, J.H.},
  \bibinfo{author}{Liu, E.H.}, \bibinfo{author}{Shim, J.},
  \bibinfo{author}{Naruse, K.}, \bibinfo{author}{Csete, M.E.},
  \bibinfo{author}{Thouless, M.}, \bibinfo{author}{Takayama, S.},
  \bibinfo{year}{2005}.
\newblock \bibinfo{title}{Fabrication of reconfigurable protein matrices by
  cracking}.
\newblock \bibinfo{journal}{Nature materials} \bibinfo{volume}{4},
  \bibinfo{pages}{403--406}.
\bibitem[{Zitnay and Weiss(2018)}]{zitnay2018load}
\bibinfo{author}{Zitnay, J.L.}, \bibinfo{author}{Weiss, J.A.},
  \bibinfo{year}{2018}.
\newblock \bibinfo{title}{Load transfer, damage, and failure in ligaments and
  tendons}.
\newblock \bibinfo{journal}{Journal of Orthopaedic Research{\textregistered}}
  \bibinfo{volume}{36}, \bibinfo{pages}{3093--3104}.

\end{thebibliography}

\end{document}